\documentclass[english,aps,notitlepage,nofootinbib,tightenlines,11pt]{revtex4-1}
\usepackage[T1]{fontenc}
\usepackage[latin9]{inputenc}
\usepackage{color}
\usepackage{amsmath}
\usepackage{graphicx}
\usepackage{esint}
\PassOptionsToPackage{normalem}{ulem}
\usepackage{ulem}

\makeatletter

\providecommand{\tabularnewline}{\\}

\pdfoutput=1
\usepackage[colorlinks=true,citecolor=blue,linkcolor=blue]{hyperref}
\usepackage{slashed}

\makeatother

\usepackage{babel}
\begin{document}

\title{Dirac neutrinos and $N_{{\rm eff}}$}

\author{Xuheng Luo$^{a,b}$, Werner Rodejohann$^b$ and Xun-Jie Xu$^b$}

\affiliation{$^{a}$University of Science and Technology of China, Hefei, Anhui
230026, China \\
$^{b}$Max-Planck-Institut f\"ur Kernphysik, Postfach 103980, D-69029
Heidelberg, Germany
}

\begin{abstract}
\noindent
If neutrinos are Dirac particles the existence of
light right-handed neutrinos $\nu_{R}$ is implied. Those would contribute to
the effective number of relativistic neutrino species $N_{{\rm eff}}$
in the early Universe. With pure standard model interactions, the
contribution is negligibly small. In the presence of new interactions, however, the contribution 
could be significantly enhanced. We consider the most general effective 
four-fermion interactions for neutrinos (scalar, pseudo-scalar, vector, axial-vector and tensor),  
and compute the contribution of right-handed neutrinos to $N_{{\rm eff}}$. 
Taking the Planck 2018 measurement of $N_{{\rm eff}}$, strong constraints on the effective four-fermion coupling are obtained, corresponding to interaction strengths of $10^{-5}\sim10^{-3}$ in units of the Fermi constant. This translates in new physics scales of up to 43 TeV and higher. Future experiments 
such as CMB-S4 can probe or exclude the existence of effective 4-neutrino operators for Dirac neutrinos. Ways to avoid this conclusion are discussed.  
\end{abstract}
\maketitle

\section{Introduction}

\noindent
One of the most important questions in neutrino physics is whether
neutrinos are Dirac or Majorana particles. The essential difference
between the two cases is that a Dirac neutrino contains two more light
degrees of freedom than a Majorana neutrino. These degrees of freedom correspond to light right-handed neutrinos ($\nu_{R}$), which are absent in the Standard Model (SM) of particle physics. While theoretically the Majorana option is generally favored, every experimental measurement so far is in agreement with 
the Dirac hypothesis \cite{Dolinski:2019nrj}. 
Indeed, many models and scenarios have been put forward that can forbid Majorana mass terms for the neutrinos and thus render neutrinos Dirac particles, see e.g.\ the review \cite{Xing:2019vks} for some references.

Even though the Dirac scenario implies the existence of $\nu_{R}$, 
it is well known that those would not contribute
significantly to the effective number of relativistic neutrino species
$N_{{\rm eff}}$ in the early Universe, provided that neutrinos only
interact as the SM predicts. With pure SM interactions, the smallness of neutrino Yukawa couplings 
means that $\nu_{R}$ would hardly couple to the SM thermal bath so that their energy 
density would be much lower than that of left-handed 
neutrinos $\nu_{L}$ --- see e.g.\ the review~\cite{Dolgov:2002wy}.

However, since the existence of tiny neutrino masses is calling for
new physics, it is reasonable to speculate that the interactions of
neutrinos may also go beyond the SM. In general, if new neutrino interactions are present, 
then right-handed neutrinos could be thermalized and contribute significantly to $N_{{\rm eff}}$.
 By requiring that the contribution does not exceed the current bound
on $N_{{\rm eff}}$, one can obtain very strong constraints on such
new interactions. This is the content of our paper. 

Already in Ref.~\cite{Masso:1994ww}, pure vector interactions of the form
$G_{V}(\overline{\nu}\gamma^{\mu}\nu)(\overline{\nu}\gamma_{\mu}\nu)$
have been considered, and   $G_{V}<3\times10^{-3}\thinspace G_{F}$, 
where $G_{F}$ is the Fermi constant,  has been derived  by simply
assuming that $\nu_{R}$ should have decoupled before the QCD phase
transition ($T\approx200$ MeV), which is roughly equivalent to $\Delta N_{{\rm eff}}={\cal O}(1)$.
Nowadays, with precision data from CMB observations, $\Delta N_{{\rm eff}}$
has been constrained more stringently. Currently the best measurement,
$N_{{\rm eff}}=2.99\pm0.17$, comes from the Planck 2018 data~\cite{Akrami:2018vks,Aghanim:2018eyx},
which is consistent with the SM prediction $N_{{\rm eff}}^{{\rm SM}}=3.045$~\cite{Mangano:2005cc,Grohs:2015tfy,deSalas:2016ztq}.
In the future, CMB Stage IV experiments (CMB-S4) are expected to reach
a precision of $\Delta N_{{\rm eff}}\sim0.03$~\cite{Abazajian:2016yjj,Abazajian:2019eic}.
A very recent study~\cite{Abazajian:2019oqj} shows that with such
 precision, the cosmological constraints on some Dirac neutrino models
such as unbroken (or adequately broken) $U(1)_{B-L}$ or neutrinophilic 2-Higgs Doublet Models could exceed most laboratory
constraints. Ref.~\cite{deSalas:2016ztq} has considered the effect
of the so-called Non-Standard Interactions (NSI) of the $V-A$
form, which have been extensively studied in the literature --- see e.g.\ the reviews~\cite{Davidson:2003ha,Ohlsson:2012kf,Farzan:2017xzy,Dev:2019anc}. The paper concluded that NSI could reduce  $N_{{\rm eff}}$, depending on the flavor structure of the new interactions, down to 3.040 
 or enhance it to 3.059. In addition
to these aforementioned scenarios, there has been a variety of other
new physics scenarios proposed in the literature~\cite{Boehm:2012gr,Kamada:2015era,Huang:2017egl,Fradette:2018hhl,Escudero:2018mvt,Escudero:2020dfa,Depta:2019lbe}
that could affect $N_{{\rm eff}}$.

In this work, we consider a set of effective four-fermion interactions 
of Dirac neutrinos with all possible Lorentz invariant forms, including
scalar, pseudo-scalar, vector, axial-vector and tensor couplings,
and study their effect on $N_{{\rm eff}}$. 
Such {\it generalized neutrino interactions} have recently been discussed intensively 
\cite{Lindner:2016wff, Rodejohann:2017vup, Kosmas:2017tsq, AristizabalSierra:2018eqm, Boehm:2018sux, Bischer:2018zcz, Xu:2019dxe, Bolton:2019wta, Chao:2019pyh, Bischer:2019ttk, Khan:2019jvr, Bolton:2020xsm,Han:2020pff}.  
Our study reveals that in this framework,
$N_{{\rm eff}}$ could be significantly enhanced from new interactions involving $\nu_R$, 
which therefore can be significantly constrained by current and future CMB experiments. Taking the
constraint on $N_{{\rm eff}}$ from the Planck 2018 data, we derive upper 
bounds on the effective four-fermion couplings of the order $10^{-5}\sim10^{-3}$ $G_{F}$, depending on the interaction 
forms. This implies that new physics up to $43$ TeV is probed. 
Future experiments such as CMB-S4 could fully exclude or probe this scenario, though there are ways to avoid this conclusion, which are discussed in this paper.\\

The paper is organized as follows: in Sec.~\ref{sec:basic}, we describe our set of new interactions, while 
in Sec.~\ref{sec:Boltzmann} we describe how those interactions enter the Boltzmann equation that describes the evolution of the right-handed neutrino density. This evolution, its effect on $N_{\rm eff}$ and the  resulting limits on the new interactions are discussed in Sec.\ \ref{sec:evol}. Conclusions are presented in   Sec.~\ref{sec:Conclusion}, and various technical details are delegated to the Appendix.

\section{General four-fermion interactions\label{sec:basic}}
\noindent
If neutrinos are Dirac particles and have new interactions beyond
the SM, the right-handed components $\nu_{R}$ could have been in
thermal equilibrium with the SM plasma.  However, observation requires that they decouple from
the SM plasma much earlier than the left-handed neutrinos $\nu_{L}$. For example, the Planck 2018 data requires that in the presence of three $\nu_R$, they should have decoupled at temperatures greater than $T>600$ MeV~\cite{Abazajian:2019oqj}. 
This implies that if $\nu_{R}$ are in thermal equilibrium with $\nu_{L}$,
then they are also in thermal equilibrium with other SM particles, 
and vice versa. Therefore, considering only interactions between $\nu_{R}$
and $\nu_{L}$ can be very representative and also greatly simplifies
the problem. 

We formulate the new interactions of Dirac neutrinos
as follows~\cite{Rodejohann:2017vup}:
\begin{equation}
{\cal L}\supset\frac{G_{F}}{\sqrt{2}}\sum_{a}\overline{\nu}\Gamma^{a}\nu\left[\overline{\nu}\Gamma^{a}(\epsilon_{a}+\tilde{\epsilon}_{a}i_{a}\gamma^{5})\nu\right],\label{eq:4f}
\end{equation}
where the index $a=(S,\thinspace P,\thinspace V,\thinspace A,\thinspace T)$
denotes scalar, pseudo-scalar, vector, axial-vector and tensor interactions, i.e.\ the 
five possible combinations of Dirac matrices that could appear between
two Dirac spinors:
\begin{equation}
\Gamma^{a}=\{I,\ i\gamma^{5},\ \gamma^{\mu},\ \gamma^{\mu}\gamma^{5},\ \sigma^{\mu\nu}\equiv\frac{i}{2}[\gamma^{\mu},\gamma^{\nu}]\}\thinspace.\label{eq:Gamma}
\end{equation}
In Eq.~(\ref{eq:4f}), we have introduced $i_{a}=i$ for $a=S,$ $P$,
$T$ and $i_{a}=1$ for $a=V,$ $A$ so that $\epsilon_{a}$ and $\tilde{\epsilon}_{a}$
are real coefficients and Eq.~(\ref{eq:4f}) is self-conjugate\footnote{Otherwise an ``${\rm h.c.}$'' term should be added and the combined
result would have the same form.}.   

In principle, one could include flavor dependence in Eq.~(\ref{eq:4f})
by adding flavor indices to $\nu$, $\overline{\nu}$, $\epsilon_{a}$
and $\tilde{\epsilon}_{a}$ --- see e.g.~\cite{Khan:2019jvr}. With 
 flavor dependence, $\nu_{R}$ of different flavors could have
different decoupling temperatures. In this work, for simplicity,  we
assume the interactions are flavor universal and flavor diagonal,
which means that the interaction in Eq.~(\ref{eq:4f}) exists for
each generation of neutrinos with the same strength.

It is useful to express $\overline{\nu}\Gamma^{a}\nu$ in terms of
chiral Dirac spinors $\nu_{L}=P_{L}\nu$ and $\nu_{R}=P_{R}\nu$, where
$P_{L/R}\equiv(1\mp\gamma^{5})/2$:
\begin{eqnarray}
\overline{\nu}\nu & = & \overline{\nu_{R}}\nu_{L}+\overline{\nu_{L}}\nu_{R}\thinspace,\label{eq:n}\\
\overline{\nu}i\gamma^{5}\nu & = & -i\overline{\nu_{R}}\nu_{L}+i\overline{\nu_{L}}\nu_{R}\thinspace,\label{eq:n-1}\\
\overline{\nu}\gamma^{\mu}\nu & = & \overline{\nu_{L}}\gamma^{\mu}\nu_{L}+\overline{\nu_{R}}\gamma^{\mu}\nu_{R}\thinspace,\label{eq:n-2}\\
\overline{\nu}\gamma^{\mu}\gamma^{5}\nu & = & -\overline{\nu_{L}}\gamma^{\mu}\nu_{L}+\overline{\nu_{R}}\gamma^{\mu}\nu_{R}\thinspace,\label{eq:n-3}\\
\overline{\nu}\sigma^{\mu\nu}\nu & = & \overline{\nu_{R}}\sigma^{\mu\nu}\nu_{L}+\overline{\nu_{L}}\sigma^{\mu\nu}\nu_{R}\thinspace,\label{eq:n-4}\\
\overline{\nu}\sigma^{\mu\nu}i\gamma^{5}\nu & = & -i\overline{\nu_{R}}\sigma^{\mu\nu}\nu_{L}+i\overline{\nu_{L}}\sigma^{\mu\nu}\nu_{R}\thinspace.\label{eq:n-5}
\end{eqnarray}
In the SM, neutrino interactions respect the $V-A$ form, which implies
that only the combination $\overline{\nu}\gamma^{\mu}\nu-\overline{\nu}\gamma^{\mu}\gamma^{5}\nu=2\,\overline{\nu_{L}}\gamma^{\mu}\nu_{L}$
is present, i.e.\ only left-handed neutrinos are involved.

Plugging Eqs.\ (\ref{eq:n})-(\ref{eq:n-5}) into Eq.\ (\ref{eq:4f}),
we obtain several interaction terms linking left- and right-handed neutrinos: 
\begin{eqnarray}
{\cal L} & \supset & G_{S}\, \overline{\nu_{L}}\nu_{R}\overline{\nu_{L}}\nu_{R}+G_{S}^{*}\, \overline{\nu_{R}}\nu_{L}\overline{\nu_{R}}\nu_{L}\nonumber \\
 & + & \tilde{G}_{S}\, \overline{\nu_{L}}\nu_{R}\overline{\nu_{R}}\nu_{L}\nonumber \\
 & + & G_{V}\, \overline{\nu_{L}}\gamma^{\mu}\nu_{L}\overline{\nu_{R}}\gamma_{\mu}\nu_{R}\nonumber \\
 & + & G_{T}\, \overline{\nu_{L}}\sigma^{\mu\nu}\nu_{R}\overline{\nu_{L}}\sigma_{\mu\nu}\nu_{R}+G_{T} ^{*}\,\overline{\nu_{R}}\sigma^{\mu\nu}\nu_{L}\overline{\nu_{R}}\sigma_{\mu\nu}\nu_{L}\,.\label{eq:b-9}
\end{eqnarray}
Here we have defined new effective 4-fermion coefficients, namely 
\begin{eqnarray}
G_{S} & = & \frac{G_{F}}{\sqrt{2}}\left(\epsilon_{S}+i\tilde{\epsilon}_{S}-\epsilon_{P}-i\tilde{\epsilon}_{P}\right),\label{eq:b-10}\\
\tilde{G}_{S} & = & \sqrt{2}G_{F}\left(\epsilon_{S}+\epsilon_{P}\right),\label{eq:b-11}\\
G_{V} & = & \sqrt{2}G_{F}(\epsilon_{V}-\epsilon_{A})\,,\label{eq:b-12}\\
G_{T} & = & \frac{G_{F}}{\sqrt{2}}\left(\epsilon_{T}+i\tilde{\epsilon}_{T}\right).\label{eq:b-13}
\end{eqnarray}
In Eq.\ (\ref{eq:b-9}), we have neglected two terms $\overline{\nu_{L}}\gamma^{\mu}\nu_{L}\overline{\nu_{L}}\gamma_{\mu}\nu_{L}$
and $\overline{\nu_{R}}\gamma^{\mu}\nu_{R}\overline{\nu_{R}}\gamma_{\mu}\nu_{R}$, 
which  cannot convert $\nu_{R}$ and $\nu_{L}$ into each other and would thus not contribute to generating  right-handed neutrino energy densities in the early Universe.

Given the four-fermion operators in Eq.~(\ref{eq:b-9}), there are
five processes relevant to the evolution of the $\nu_{R}$ abundance
($\overline{\nu_{R}}$ has exactly the same thermal dynamics as $\nu_{R}$)
in the early Universe:
\begin{eqnarray}
\nu_{R}+\nu_{R} & \leftrightarrow & \nu_{L}+\nu_{L}\thinspace,\label{eq:b-14}\\
\nu_{R}+\overline{\nu_{R}} & \leftrightarrow & \nu_{L}+\overline{\nu_{L}}\thinspace,\label{eq:b-15}\\
\nu_{R}+\nu_{L} & \leftrightarrow & \nu_{R}+\nu_{L}\thinspace,\label{eq:b-16}\\
\nu_{R}+\overline{\nu_{L}} & \leftrightarrow & \nu_{R}+\overline{\nu_{L}}\thinspace,\\
\nu_{R}+\overline{\nu_{L}} & \leftrightarrow & \overline{\nu_{R}}+\nu_{L}\thinspace.\label{eq:b-17}
\end{eqnarray}
The scattering matrix elements of the above processes are computed
in Appendix~\ref{sec:M2}. The result is summarized in Tab.~\ref{tab:t}. Note that when one of the above processes (\ref{eq:b-14})-(\ref{eq:b-17}) is present, right-handed neutrinos are automatically generated in the early Universe. For instance, in the presence of Eq.\ (\ref{eq:b-16}), which on its own would not generate right-handed neutrinos without an initial population, the process in Eq.\ (\ref{eq:b-15}) necessarily exists because the same couplings are involved. Hence, right-handed neutrinos are produced. 

\begin{table*}
\caption{\label{tab:t}Processes that involve $\nu_{R}$ as initial or final
states and the corresponding scattering matrix elements $|{\cal M}|^{2}$,
assuming the presence of all  terms in Eq.~(\ref{eq:b-9}). Note
that when used in phase space integrals containing identical particles,
the matrix elements need to be multiplied by an additional symmetry
factor $S$, see Eq.~(\ref{eq:b-1-1-1}),  which is not  included in this table.}

\begin{ruledtabular}
\begin{tabular}{cccc}
 & process & $|{\cal M}|^{2}$ & \tabularnewline
\hline 
 & $\nu_{R}(p_{1})+\nu_{R}(p_{2})\leftrightarrow\nu_{L}(p_{3})+\nu_{L}(p_{4})$ & $16|G_{S}-12G_{T}|^{2}(p_{1}\cdot p_{2})(p_{3}\cdot p_{4})$ & \tabularnewline
 & $\nu_{R}(p_{1})+\overline{\nu_{R}}(p_{2})\leftrightarrow\nu_{L}(p_{3})+\overline{\nu_{L}}(p_{4})$ & $4|\tilde{G}_{S}-2G_{V}|^{2}(p_{1}\cdot p_{3})(p_{2}\cdot p_{4})$ & \tabularnewline
 & $\nu_{R}(p_{1})+\nu_{L}(p_{2})\leftrightarrow\nu_{R}(p_{3})+\nu_{L}(p_{4})$ & $4|\tilde{G}_{S}-2G_{V}|^{2}(p_{1}\cdot p_{4})(p_{3}\cdot p_{2})$ & \tabularnewline
 & $\nu_{R}(p_{1})+\overline{\nu_{L}}(p_{2})\leftrightarrow\nu_{R}(p_{3})+\overline{\nu_{L}}(p_{4})$ & $4|\tilde{G}_{S}-2G_{V}|^{2}(p_{1}\cdot p_{2})(p_{3}\cdot p_{4})$ & \tabularnewline
 & $\nu_{R}(p_{1})+\overline{\nu_{L}}(p_{2})\leftrightarrow\overline{\nu_{R}}(p_{3})+\nu_{L}(p_{4})$ & $16|G_{S}-12G_{T}|^{2}(p_{1}\cdot p_{3})(p_{2}\cdot p_{4})$ & \tabularnewline
\end{tabular}\end{ruledtabular}

\end{table*}

\section{Boltzmann equation\label{sec:Boltzmann}}
\noindent 
Recall that for a spatially homogeneous and isotropic Universe, we have 
\begin{equation}
\dot{\rho}_{{\rm tot}}+3H(\rho_{{\rm tot}}+P_{{\rm tot}})=0,\label{eq:b-29}
\end{equation}
where $H^2 = (\dot a/a)^2 = \frac{8\pi}{3m_{{\rm Pl}}^{2}}\rho_{{\rm tot}}$ is the Hubble parameter and 
$\rho_{{\rm tot}}$ and $P_{{\rm tot}}$ are the total energy density and pressure, respectively.
Since we introduce $\nu_{R}$ to the SM, $\rho_{{\rm tot}}$ and $P_{{\rm tot}}$
can be decomposed as
\begin{eqnarray*}
\rho_{{\rm tot}} & = & \rho_{{\rm SM}}+\rho_{\nu_{R}},\\
P_{{\rm tot}} & = & P_{{\rm SM}}+P_{\nu_{R}},
\end{eqnarray*}
where the subscripts "SM" and "$\nu_{R}$" denote the contributions from SM 
particles and from $\nu_{R}$, respectively. The latter in general have a temperature $T_{\nu_{R}}$ that is different from the one of the SM particles $T_{\rm SM}$, which we can consider to be the same for all SM particles, see the discussion after Eq.\ (\ref{eq:b-36}).
Without any interactions between SM particles
and $\nu_{R}$, Eq.~(\ref{eq:b-29}) could be applied to $\rho_{{\rm SM}}$
and $\rho_{\nu_{R}}$ individually, with the subscript "tot" replaced
by ''SM" and "$\nu_{R}$". However, in the presence of $\nu_{R}$-SM interactions, 
there is energy transfer between the two components, which leads to
the following evolution equations for $\rho_{{\rm SM}}$ and $\rho_{\nu_{R}}$:
\begin{eqnarray}
\dot{\rho}_{{\rm SM}}+3H(\rho_{{\rm SM}}+P_{{\rm SM}}) & = & -C_{\nu_{R}}^{(\rho)},\label{eq:b-30}\\
\dot{\rho}_{\nu_{R}}+3H(\rho_{\nu_{R}}+P_{\nu_{R}}) & = & C_{\nu_{R}}^{(\rho)},\label{eq:b-31}
\end{eqnarray}
where $C_{\nu_{R}}^{(\rho)}$, known as a collision term, can be physically
interpreted as  the energy transfer rate from SM particles to $\nu_{R}$.
Taking the sum of Eqs.~(\ref{eq:b-30}) and
(\ref{eq:b-31}) yields again, as it should,  Eq.~(\ref{eq:b-29}).
The explicit form of $C_{\nu_{R}}^{(\rho)}$ is derived from Boltzmann
equations (see Appendix~\ref{sec:thermaldy}), and given as follows:
\begin{eqnarray}
C_{\nu_{R}}^{(\rho)} & = & -N_{\nu_{R}}\int E_{1}d\Pi_{1}d\Pi_{2}d\Pi_{3}d\Pi_{4}(2\pi)^{4}\delta^{4}(p_{1}+p_{2}-p_{3}-p_{4})\nonumber \\
 &  & \times S\left[|{\cal M}|_{1+2\rightarrow3+4}^{2}f_{1}f_{2}(1-f_{3})(1-f_{4})\right.\nonumber \\
 &  & \ \ \ \ \left.-|{\cal M}|_{3+4\rightarrow1+2}^{2}f_{3}f_{4}(1-f_{1})(1-f_{2})\right],\label{eq:b-1-1-1}
\end{eqnarray}
\begin{equation}
d\Pi_{i}\equiv\frac{g_{i}}{(2\pi)^{3}}\frac{d^{3}p_{i}}{2E_{i}},\ \ f_{i}\equiv\frac{1}{\exp\left(\frac{E_{i}}{T_{i}}\right)+1},\ (i=1,\thinspace2,\thinspace3,\thinspace4).\label{eq:b-32}
\end{equation}
Here $1+2\rightarrow3+4$ represents the processes listed in Tab.~\ref{tab:t}
and $3+4\rightarrow1+2$ represents the inverse processes; $g_{i}$, $E_{i}$
and $T_{i}$ are the number of internal degrees of freedom, energy,
and temperature of particle $i$. Without loss of generality, we always
assign 1 to $\nu_{R}$ while $2$, $3$, and $4$ may be assigned
to any of $\nu_{R}$, $\overline{\nu_{R}}$, $\nu_{L}$, and $\overline{\nu_{L}}$,
depending on the processes taken from Tab.~\ref{tab:t};  $S$ is
a symmetry factor related to the number of identical particles in
the initial/final states and $N_{\nu_{R}}$ is the number of right-handed
neutrinos, both to be explained  in detail later, see Eq.\ (\ref{eq:b-34}).

The collision term $C_{\nu_R}^{(\rho)}$ in Eq.~(\ref{eq:b-1-1-1})
is a $12$-dimensional integral, hence the numerical evaluation can be
very time-consuming. Fortunately, by applying a technique developed
in Refs.~\cite{Hannestad:1995rs,Dolgov:1997mb}\footnote{For practical use, we refer the readers to Appendix A of Ref.~\cite{Dolgov:1997mb}
and Appendix D in Ref.~\cite{Fradette:2018hhl}.}, we can drastically reduce it  to a 3-dimensional integral.  For
massless particles obeying Maxwell-Boltzmann statistics, the 3-dimensional
integral can be further integrated analytically, giving a purely analytical
result. Neutrinos in our work can  be treated as massless
particles but they obey Fermi-Dirac statistics. Nevertheless, by expanding
the Fermi-Dirac distributions in terms of $\exp\left(-E/T\right)$
with Maxwell-Boltzmann distributions as leading order approximation, one can
obtain a good approximation~\cite{Dolgov:2002wy}. A recent numerical
study in Ref.~\cite{Escudero:2020dfa} shows that the difference
between Fermi-Dirac and Maxwell-Boltzmann statistics can be accounted for 
by multiplying the collision terms 
with a factor of $1-\delta_{{\rm FD}}$ with $\delta_{{\rm FD}}={\cal O}(0.1)$.
In Appendix~\ref{sec:coll}, we use the technique of Ref.~\cite{Dolgov:1997mb}
to reduce the dimension of the integral and integrate the 3-dimensional 
integral analytically for Maxwell-Boltzmann statistics. The result is given in the third column of 
 Tab.~\ref{tab:t-1}. 
The Fermi-Dirac correction factor $1-\delta_{{\rm FD}}$ then can be obtained by solving the 
integral numerically, and comparing to the analytical Maxwell-Boltzmann result.  
We performed this in a way similar to that  used  in Ref.~\cite{Escudero:2020dfa}, and our results for the five relevant scattering 
processes are summarized in Tab.~\ref{tab:t-1}. We have checked
that when our code is applied to $\nu_{L}$ decoupling in  standard
cosmology (i.e., $\nu_{L}$-$e$ scattering), the Fermi-Dirac correction 
factors reported in Ref.~\cite{Escudero:2020dfa} can be reproduced.

\begin{table*}
\caption{\label{tab:t-1}Collision terms $C_{\nu_{R}}^{(\rho)}$ computed for
all the relevant processes including left- and right-handed neutrinos. 
In the third column, the $C_{\nu_{R}}^{(\rho)}$
are computed analytically from Maxwell-Boltzmann (MB) statistics. When used for
neutrinos, the $C_{\nu_{R}}^{(\rho)}$ should be multiplied with 
Fermi-Dirac correction factors $1-\delta_{{\rm FD}}$ in the last
column.}

\begin{ruledtabular}
\begin{tabular}{cccccc}
 & process & $S$ & $C_{\nu_{R}}^{(\rho)}$ from MB statistics & $1-\delta_{{\rm FD}}$ & \tabularnewline
\hline 
 & $\nu_{R}(p_{1})+\nu_{R}(p_{2})\leftrightarrow\nu_{L}(p_{3})+\nu_{L}(p_{4})$ & $\frac{2}{2!2!}$ &  $\frac{12}{\pi^{5}}|G_{S}-12G_{T}|^{2}N_{\nu_{R}}\left(T_{{\rm SM}}^{9}-T_{\nu_{R}}^{9}\right)$ & 0.8840 & \tabularnewline
 & $\nu_{R}(p_{1})+\overline{\nu_{R}}(p_{2})\leftrightarrow\nu_{L}(p_{3})+\overline{\nu_{L}}(p_{4})$ & 1 & $\frac{2}{\pi^{5}}|\tilde{G}_{S}-2G_{V}|^{2}N_{\nu_{R}}\left(T_{{\rm SM}}^{9}-T_{\nu_{R}}^{9}\right)$ & 0.8841 & \tabularnewline
 & $\nu_{R}(p_{1})+\nu_{L}(p_{2})\leftrightarrow\nu_{R}(p_{3})+\nu_{L}(p_{4})$ & 1 & $\frac{1}{2\pi^{5}}|\tilde{G}_{S}-2G_{V}|^{2}N_{\nu_{R}}T_{{\rm SM}}^{4}T_{\nu_{R}}^{4}\left(T_{{\rm SM}}-T_{\nu_{R}}\right)$ & 0.8518 & \tabularnewline
 & $\nu_{R}(p_{1})+\overline{\nu_{L}}(p_{2})\leftrightarrow\nu_{R}(p_{3})+\overline{\nu_{L}}(p_{4})$ & 1 & $\frac{3}{\pi^{5}}|\tilde{G}_{S}-2G_{V}|^{2}N_{\nu_{R}}T_{{\rm SM}}^{4}T_{\nu_{R}}^{4}\left(T_{{\rm SM}}-T_{\nu_{R}}\right)$ & 0.8249 & \tabularnewline
 & $\nu_{R}(p_{1})+\overline{\nu_{L}}(p_{2})\leftrightarrow\overline{\nu_{R}}(p_{3})+\nu_{L}(p_{4})$ & 1 & $\frac{6}{\pi^{5}}|G_{S}-12G_{T}|^{2}N_{\nu_{R}}T_{{\rm SM}}^{4}T_{\nu_{R}}^{4}\left(T_{{\rm SM}}-T_{\nu_{R}}\right)$ & 0.8118 & \tabularnewline
\end{tabular}\end{ruledtabular}

\end{table*}

Eqs.~(\ref{eq:b-30}) and (\ref{eq:b-31}) are the main expressions 
of this work, which allow to determine the evolution of $\nu_{R}$
energy density  in the early Universe. 
Three comments are given as follows:

First of all, for $f_{i}$ in Eq.~(\ref{eq:b-32}), we have assumed
that there is no spectral distortion of neutrino energy distributions
when they decouple from the SM plasma. This is only true if the effective
degrees of freedom of the SM are  constant during the decoupling (for
instantaneous decoupling there would be no spectral distortion). In reality, this never strictly holds but can be taken as a good approximation. 
The well-studied SM neutrino decoupling may help to understand the
magnitude of such spectral distortions. The SM neutrinos $\nu_{L}$ 
decouple from the SM plasma at $T$ around $1$ or $2$ MeV, followed by
electron-position annihilation at $T\sim0.5$ MeV. Note that none
of these processes is instantaneous. Hence the annihilation is expected
to slightly heat up the tail of $f_{\nu_L}(E)$, which leads to a known spectral
distortion of neutrinos. Numerical calculations find that the distortion
is about \cite{Dolgov:2002wy} 
\begin{equation}
\delta f_{\nu_L}(E)/f_{\nu_L}(E)\sim3\times10^{-4}\, \frac{E}{T}\left(\frac{11}{4}\frac{E}{T}-3\right).\label{eq:b-33}
\end{equation}
For $E\sim T$, Eq.~(\ref{eq:b-33}) implies a 0.01\,\% distortion. Therefore,
in our study of $\nu_{R}$ decoupling, the spectral distortion should
be negligible.

The second comment concerns the symmetry factor $S$. 
Here we prefer to include it not in $|{\cal M}|^{2}$, so that the matrix elements 
respect crossing symmetry which is used in our calculation --- see
Appendix~\ref{sec:M2}. In the absence of identical particles, $S=1$.
In general, when there are $n$ identical particles in the initial
or final states, $S$ should be multiplied by a factor of $\frac{1}{n!}$.
If it happens that the particle is the $\psi$ in $C_\psi^{(\rho)}$ 
(in this case, $\nu_{R}$) 
is among the identical particles, one needs to further multiply $S$
by $n$.\footnote{See also the discussion below Eq.~(71) in Ref.~\cite{Dolgov:2002wy}.}
Therefore, taking $\nu_{R}+\nu_{R}\rightarrow\nu_{L}+\nu_{L}$ as an 
example, we have $S=\frac{1}{2!}\times\frac{1}{2!}\times2=1/2$, while
for $\nu_{R}+\overline{\nu_{L}}\rightarrow\overline{\nu_{R}}+\nu_{L}$, 
which is related to the former by crossing symmetry, we put $S=1$.

The last comment is on the internal degrees of freedom and differences
between $\nu$ and $\overline{\nu}$. 
In standard cosmology, $\nu_{L}$
and $\overline{\nu_{L}}$ are often treated as the same relativistic
species with two internal degrees of freedom, i.e.\ $g_{i}=2$ in Eq.~(\ref{eq:b-32}). 
In this convention, $\rho_{\nu_L} $ stands for the total energy density of both 
$\nu_L$ and $\overline{\nu_L}$. 
In principle, one could adopt a similar
approach for $\nu_{R}$ and $\overline{\nu_{R}}$. However, when using
the results in Tab.~\ref{tab:t}, one may notice that  some operators
(e.g., $G_{S}\, \overline{\nu_{L}}\nu_{R}\overline{\nu_{L}}\nu_{R}$)
can produce $\nu_{R}+\overline{\nu_{L}}\leftrightarrow\overline{\nu_{R}}+\nu_{L}$,
but not $\nu_{R}+\nu_{L}\leftrightarrow\nu_{R}+\nu_{L}$. A simple
way to cope with such a difference is that we treat $\nu_{L}$ and
$\overline{\nu_{L}}$ as two different species with the same temperature
(physically they are different as  they have opposite lepton numbers).
Likewise, $\nu_{R}$ and $\overline{\nu_{R}}$ are also treated as
two different species. This implies that  we treat $\nu_{L}$, $\overline{\nu_{L}}$,
$\nu_{R}$, and $\overline{\nu_{R}}$ as four different species. Each
of them has only one internal degree of freedom, i.e., we take $g_{i}=1$
in Eq.~(\ref{eq:b-32}). In this approach, Eq.~(\ref{eq:b-31}) would
actually split into two equations for $\nu_{R}$ and $\overline{\nu_{R}}$,
with independent energy densities ($\rho_{\nu_{R}}$ and $\rho_{\overline{\nu_{R}}}$),
pressures ($P_{\nu_{R}}$ and $P_{\overline{\nu_{R}}}$), and collision
terms ($C_{\nu_{R}}^{(\rho)}$ and $C_{\overline{\nu_{R}}}^{(\rho)}$),
respectively. Since $\nu_{R}$ and $\overline{\nu_{R}}$ have the
same temperature ($T_{\nu_{R}}=T_{\overline{\nu_{R}}}$), one would
have $\rho_{\nu_{R}}=\rho_{\overline{\nu_{R}}}$, $P_{\nu_{R}}=P_{\overline{\nu_{R}}}$,
and $C_{\nu_{R}}^{(\rho)}=C_{\overline{\nu_{R}}}^{(\rho)}$, which
would allow us to recombine them. Therefore, even though
we conceptually split them, we do not need to do so explicitly. We
can simply take them as different species, then focus on $\nu_{R}$
to compute the phase space integral, and eventually multiply the result
by the number of different species. In this way, it is also straightforward 
to include three generations of neutrinos ($N_{\nu}=3$) in our analyses.
In conclusion, for $g_{i}$ in Eq.~(\ref{eq:b-32}), $\rho_{\nu_{R}}$
and $P_{\nu_{R}}$ in Eq.~(\ref{eq:b-31}), and $N_{\nu_{R}}$ in
Eq.~(\ref{eq:b-1-1-1}), we take
\begin{equation}
g_{i}=1,\ \ \rho_{\nu_{R}}=2\times N_{\nu}\times\frac{7}{8}\times\frac{\pi^{2}}{30}T_{\nu_{R}}^{4},\ \ P_{\nu_{R}}=2\times N_{\nu}\times\frac{7}{8}\times\frac{\pi^{2}}{90}T_{\nu_{R}}^{4},\ \ N_{\nu_{R}}=2\times N_{\nu}.\label{eq:b-34}
\end{equation}
Here $\frac{7}{8}\times\frac{\pi^{2}}{30}T_{\nu_{R}}^{4}$ and $\frac{7}{8}\times\frac{\pi^{2}}{90}T_{\nu_{R}}^{4}$
are contributions from one fermionic degree of freedom --- see Eqs.~(\ref{eq:b-4-2})
and (\ref{eq:b-6-2}) for a brief review of the related thermodynamics. 
The SM energy density and pressure can be computed from
\begin{equation}
\rho_{{\rm SM}}=\frac{\pi^{2}}{30}g_{\star}^{(\rho)}T_{{\rm SM}}^{4},\ \ P_{{\rm SM}}=\frac{\pi^{2}}{90}g_{\star}^{(P)}T_{{\rm SM}}^{4},\label{eq:rhoPSM}
\end{equation}
where $g_{\star}^{(\rho)}$ and $g_{\star}^{(P)}$ are effective degrees
of freedom of the SM. The SM contains $3\times3\times4$ quarks, $3\times2$
charged leptons, $3$ left-handed neutrinos, $8+3+1$ gauge bosons,
and one Higgs doublet. Therefore, at a sufficiently high temperature
($T_{{\rm SM}}\gg$ any SM particle mass), we have $g_{\star}^{(\rho)}=g_{\star}^{(P)}=(36+6+3)\times2\times7/8+12\times2+4=106.75$.
For lower temperatures, the calculations of $g_{\star}^{(\rho)}$
and $g_{\star}^{(P)}$ can be very involved. Note that, in general,
$g_{\star}^{(\rho)}\neq g_{\star}^{(P)}$ and $dg_{\star}^{(\rho,P)}/dT_{{\rm SM}}\neq0$. 
 We refer to Ref.~\cite{Husdal:2016haj} for the latest results
of $g_{\star}^{(\rho)}$ and $g_{\star}^{(P)}$, which are used in
our numerical calculations.

\section{Evolution of $\nu_{R}$ abundance in the early Universe\label{sec:evol}}
\noindent 
To understand the behavior of right-handed neutrinos in the presence of new interactions, we start with the ratio of Eqs.~(\ref{eq:b-30}) and (\ref{eq:b-31}):
\begin{equation}
\frac{d\rho_{\nu_{R}}}{d\rho_{{\rm SM}}}=\frac{3H(\rho_{\nu_{R}}+P_{\nu_{R}})-C_{\nu_{R}}^{(\rho)}}{3H(\rho_{{\rm SM}}+P_{{\rm SM}})+C_{\nu_{R}}^{(\rho)}}.\label{eq:b-35}
\end{equation}
Since $\rho_{\nu_{R}}$ and $\rho_{{\rm SM}}$ are functions of $T_{\nu_{R}}$
and $T_{{\rm SM}}$, respectively, we can replace $d\rho_{\nu_{R}}\rightarrow\frac{\partial\rho_{\nu_{R}}}{\partial T_{\nu_{R}}}dT_{\nu_{R}}$
and $d\rho_{{\rm SM}}\rightarrow\frac{\partial\rho_{{\rm SM}}}{\partial T_{{\rm SM}}}dT_{{\rm SM}}$
in Eq.~(\ref{eq:b-35}), leading to:
\begin{equation}
\frac{dT_{\nu_{R}}}{dT_{{\rm SM}}}=\frac{3H(\rho_{\nu_{R}}+P_{\nu_{R}})-C_{\nu_{R}}^{(\rho)}}{3H(\rho_{{\rm SM}}+P_{{\rm SM}})+C_{\nu_{R}}^{(\rho)}}\frac{\partial\rho_{{\rm SM}}}{\partial T_{{\rm SM}}}\left(\frac{\partial\rho_{\nu_{R}}}{\partial T_{\nu_{R}}}\right)^{-1}.\label{eq:b-36}
\end{equation}
Note that all the quantities on the right-hand side of Eq.~(\ref{eq:b-36})
are essentially functions of $T_{\nu_{R}}$ and $T_{{\rm SM}}$. Regarding
$T_{\nu_{R}}$ as a function of $T_{{\rm SM}}$, the function $T_{\nu_{R}}(T_{{\rm SM}})$
is fully determined by the differential equation (\ref{eq:b-36}).

We use Eq.~(\ref{eq:b-36}) to compute the evolution of $T_{\nu_{R}}$ (starting with $T_{\nu_{R}}=T_{{\rm SM}}$) down to a few MeV before $\nu_{L}$ decouple. After $\nu_{L}$ decouple,
the SM plasma itself splits into two decoupled components: (i) a photon 
and electron-position plasma with a temperature denoted as $T_{\gamma}$,
and (ii) left-handed neutrinos\footnote{Strictly speaking, the decoupling of $\nu_{L}$ is flavor dependent, which however does not affect our discussions and analyses below. } with a temperature $T_{\nu_{L}}$. In principle, we would need to split Eq.~(\ref{eq:b-30}) into two
equations to appropriately describe the evolution of the now two SM contributions. However, this is
not necessary because at this stage, $\nu_{R}$ must have been completely
decoupled otherwise their contributions to $N_{{\rm eff}}$ would
obviously be too large. In other words, $C_{\nu_{R}}^{(\rho)}$ at 
this temperature is extremely small compared to the Hubble expansion
term in Eq.~(\ref{eq:b-31}), so we can safely turn it off. In this case Eq.~(\ref{eq:b-31})
simply implies $d\rho_{\nu_{R}}+4\rho_{\nu_{R}}d\ln a=0$, or, in a more
familiar form, $T_{\nu_{R}}\propto a^{-1}$. All free-streaming
relativistic species have the same $T(a)$ dependence. 

Under this approximation, the final temperature of $\nu_{R}$ can
be determined by 
\begin{equation}
  \frac{T_{\nu_{R},0}}{T_{\nu_{R},10}}=\frac{a_{0}^{-1}}{a_{10}^{-1}}=\left(\frac{4}{11}\right)^{\!1/3}\frac{T_{\gamma,0}}{T_{\gamma,10}},
\label{eq:d-15}
\end{equation}
where the subscript ``$0$'' denotes any time after electron-position
annihilation, and the subscript ``10'' denotes the time when $T_{{\rm SM}}=10$
MeV. 
The first identity in Eq.~(\ref{eq:d-15}) follows from the aforementioned relation  $T_{\nu_{R}}\propto a^{-1}$ and  the second identity is the result of the expression $a_{0}^{3}T_{\gamma,0}^{3}=\frac{11}{4}a_{10}^{3}T_{\gamma,10}^{3}$ in standard cosmology. It can be derived from entropy conservation:
$g_{\star,10}^{(s)}T_{\gamma,10}^{3}a_{10}^{3}=(2T_{\gamma,0}^{3}+6\times\frac{7}{8}T_{\nu_{L},0}^{3})a_{0}^{3}$, 
where $g_{\star,10}^{(s)}=4\times7/8+3\times2\times7/8+2=10.75$ ($4$
from electrons, $3\times2$ from neutrinos, and $2$ from photons) are the relativistic degrees of freedom of the SM at 10 MeV and $T_{\nu_{L},0}/T_{\gamma,0}=(4/11)^{1/3}$ is the ratio
of the final temperatures of left-handed neutrinos and photons. 
Here we choose 10 MeV as a benchmark temperature because at this temperature all other SM particles can be safely neglected, and $\nu_L$ are still tightly coupled to electrons. These conditions allow us to compute $g_{\star,10}^{(s)}$ by simply counting the numbers of fermions and bosons.

According to the definition of $N_{{\rm eff}}$~\cite{Mangano:2005cc,deSalas:2016ztq},
the contribution of $\nu_{R}$ to $N_{{\rm eff}}$ is given by:
\begin{equation}
\Delta N_{{\rm eff}}=\frac{8}{7}\left(\frac{11}{4}\right)^{\!4/3}\frac{\rho_{\nu_{R},0}}{\rho_{\gamma,0}}.\label{eq:d-16}
\end{equation}
Using Eqs.~(\ref{eq:b-34}), (\ref{eq:rhoPSM}) and (\ref{eq:d-15}),
we have:
\begin{equation}
\Delta N_{{\rm eff}}=N_{\nu}\left(\frac{11}{4}\right)^{\!4/3}\frac{T_{\nu_{R,0}}^{4}}{T_{\gamma,0}^{4}}=N_{\nu}\left(\frac{T_{\nu_{R},10}}{T_{\gamma,10}}\right)^{4}.\label{eq:d-17}
\end{equation}
Therefore, to obtain $\Delta N_{{\rm eff}}$, we only need to solve
Eq.~(\ref{eq:b-36}) to obtain the temperature ratio at 10 MeV, which
according to Eq.~(\ref{eq:d-17}) immediately gives $\Delta N_{{\rm eff}}$.

\begin{figure}
\centering

\includegraphics[width=12cm]{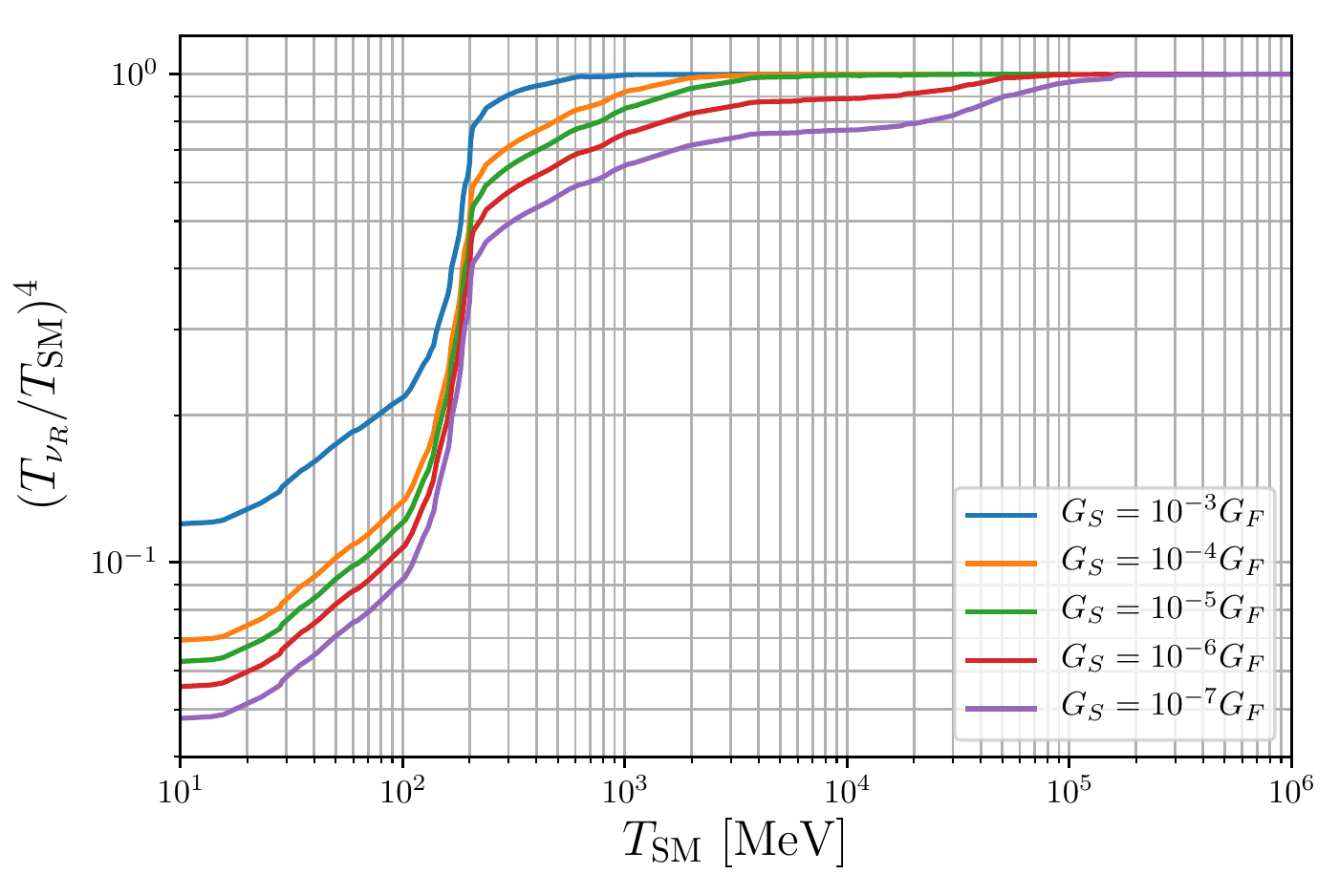}

\caption{Temperature evolution of right-handed neutrinos for different new interactions, see 
Eq.\ (\ref{eq:b-11}).\label{fig:T_R} }
\end{figure}

With Eqs.~(\ref{eq:b-34}) and (\ref{eq:rhoPSM}) and
the results in Tab.~\ref{tab:t-1}, it is straightforward to solve
 Eq.~(\ref{eq:b-36}). In Fig.~\ref{fig:T_R}, we present some solutions
for $G_{S}$ (see Eq.\ (\ref{eq:b-10})) ranging from $10^{-3}\ G_{F}$ to $10^{-7}\ G_{F}$,
assuming other interactions ($\tilde{G}_{S}$, $G_{V}$, $G_{T}$)
are absent\footnote{In Fig.~\ref{fig:T_R}, the initial value was set by $T_{\nu_{R}}=T_{{\rm SM}}$. If it had been set to zero, the curves would quickly
reache the SM temperature and the result is not changed. }. For other interactions, the curves are very similar. As
we have just mentioned, we only solve the evolution equation down
to 10 MeV, and the temperature ratio $T_{\nu_R}/T_{\gamma}$
can be directly used in Eq.~(\ref{eq:d-17})
to obtain $\Delta N_{{\rm eff}}$. For example, the curve of $G_{S}=10^{-4}\ G_{F}$
ends at $0.069$, which implies that $\Delta N_{{\rm eff}}=3\times0.069=0.21$.
For larger $G_{S}$, right-handed neutrinos decouple at lower temperatures,
leading to higher values of $T_{\nu_{R}}/T_{{\rm SM}}$ at the end, 
hence implying larger contributions to $N_{{\rm eff}}$. \\

Let us support the numerical calculation with analytical considerations.
Although the decoupling process is not instantaneous, one can nevertheless
define a decoupling temperature $T_{{\rm dec}}$ from the condition 
\begin{equation}
H\sim-\left.\frac{\partial C_{\nu_{R}}^{(\rho)}}{\partial\rho_{\nu_{R}}}\right|_{T_{\nu_{R}}=T_{{\rm SM}}\equiv T_{{\rm dec}}}.\label{eq:d-19}
\end{equation}
Using the results in Tab.~\ref{tab:t-1}, we get:
\begin{equation}
H\sim T_{{\rm dec}}^{5}G_{{\rm eff}}^{2},\label{eq:d-20}
\end{equation}
where $G_{{\rm eff}}$ is some combination of the coupling constants defined in Eq.\ (\ref{eq:b-9}), which can be estimated to be $G_{{\rm eff}}^2 = {\cal O}(0.1) G_X^2$. To be precise, we obtain by analytically evaluating the right-hand side of Eq.~(\ref{eq:d-19}) with the numerical values of $1-\delta_{\rm FD}$ in Tab.~\ref{tab:t-1} that 
\begin{equation}
G_{{\rm eff}}^{2}\equiv0.28(G_{S}-12G_{T})^{2}+0.053(\tilde{G}_{S}-2G_{V})^{2}.\label{eq:d-21}
\end{equation}
Combining Eq.~(\ref{eq:d-20}) with Eqs.~(\ref{eq:b-34}) 
and (\ref{eq:rhoPSM}), we can solve for $T_{{\rm dec}}$: 
\begin{equation}
T_{{\rm dec}}\sim1.2\times\frac{\left(g_{*}^{(\rho)}+7N_{\nu}/4\right)^{\!1/6}}{\left(G_{{\rm eff}}^{2}m_{{\rm pl}}\right)^{\!1/3}}.\label{eq:d-22}
\end{equation}
Here $g_{*}^{(\rho)}$ is a temperature-dependent quantity (106.75
at $T_{{\rm SM}}\gg100$ GeV, and 10.75 at $T_{{\rm SM}}=10$ MeV)
but the variation can be ignored due to the suppression by the exponent 1/6 (e.g.,
$10^{1/6}\approx1.5$ and $100^{1/6}\approx2.2$ are of the same order
of magnitude).  Taking $(g_{*}^{(\rho)}+7N_{\nu}/4)^{1/6}\approx2$,
we can reformulate Eq.~(\ref{eq:d-22}) as 
\begin{equation}
T_{{\rm dec}}\sim2\ {\rm MeV}\times\left(\frac{G_{{\rm eff}}}{G_{F}}\right)^{\!-2/3}.\label{eq:d-23}
\end{equation}
Taking for example $G_{S}\approx10^{-7}\ G_{F}$,  we have $G_{{\rm eff}}\approx5.3\times10^{-8}\ G_{F}$
and $T_{{\rm dec}}\sim1.4\times10^{5}$ MeV, which is qualitatively
consistent with the purple curve presented in Fig.~\ref{fig:T_R}.

If the decoupling temperature is sufficiently high, the final temperature
can be computed from entropy conservation. The entropy densities of
$\nu_{R}$ and SM are (see Appendix~\ref{sec:thermaldy}): 
\begin{equation}
s_{\nu_{R}}=\frac{2\pi^{2}}{45}N_{\nu_{R}}\frac{7}{8}T_{\nu_{R}}^{3},\ \ s_{{\rm SM}}=\frac{2\pi^{2}}{45}T_{{\rm SM}}^{3}g_{\star}^{(s)}.\label{eq:d-28}
\end{equation}
After  decoupling, the entropy of $\nu_{R}$ and the entropy of
the SM in a co-moving volume are conserved separately, i.e., $s_{\nu_{R}}a^{3}$
and $s_{{\rm SM}}a^{3}$ remain constant. This gives
\begin{equation}
\begin{cases}
T_{{\rm dec}}^{3}a_{{\rm dec}}^{3}=T_{\nu_{R},10}^{3}a_{10}^{3}\\
T_{{\rm dec}}^{3}a_{{\rm dec}}^{3}g_{\star,{\rm dec}}^{(s)}=T_{{\rm SM},10}^{3}a_{10}^{3}g_{\star,10}^{(s)}
\end{cases},\label{eq:d-24}
\end{equation}
where the subscripts "dec" and "10" denote the moments of
$\nu_{R}$ decoupling and of $T_{{\rm SM}}=10$ MeV, respectively.
The ratio of the two expressions in Eq.~(\ref{eq:d-24}) results in:
\begin{equation}
\frac{T_{\nu_{R},10}^{3}}{T_{{\rm SM},10}^{3}}=\frac{g_{\star,10}^{(s)}}{g_{\star}^{(s)}(T_{{\rm dec}})}.\label{eq:d-25}
\end{equation}
Taking $g_{\star,10}^{(s)}=4\times7/8+3\times2\times7/8+2=10.75$
($4$ from electrons, $3\times2$ from neutrinos, $2$ from photons)
and $g_{\star,{\rm dec}}^{(s)}=106.75$ (the maximal value in the
SM), we get 
\begin{equation}
T_{\nu_{R},10}/T_{{\rm SM},10}=0.465,\ \ T_{\nu_{R},10}^{4}/T_{{\rm SM},10}^{4}=0.0468.\label{eq:d-26}
\end{equation}
This roughly matches the end of the lowest violett curve in Fig.~\ref{fig:T_R}. 
Therefore, if all  three $\nu_{R}$ decouple at a temperature much
higher than the electroweak scale, according to Eqs.~(\ref{eq:d-17})
and (\ref{eq:d-26}), one would get $\Delta N_{{\rm eff}}=3\times0.0468=0.14$. 
Lower decoupling temperatures would lead to larger $\Delta N_{{\rm eff}}$, as demonstrated by the 
result of Ref.\ \cite{Abazajian:2019oqj}: Planck 2018 data implies that three right-handed neutrinos should have decoupled at temperatures larger than 600 MeV.

\begin{figure}
\centering

\includegraphics[width=12cm]{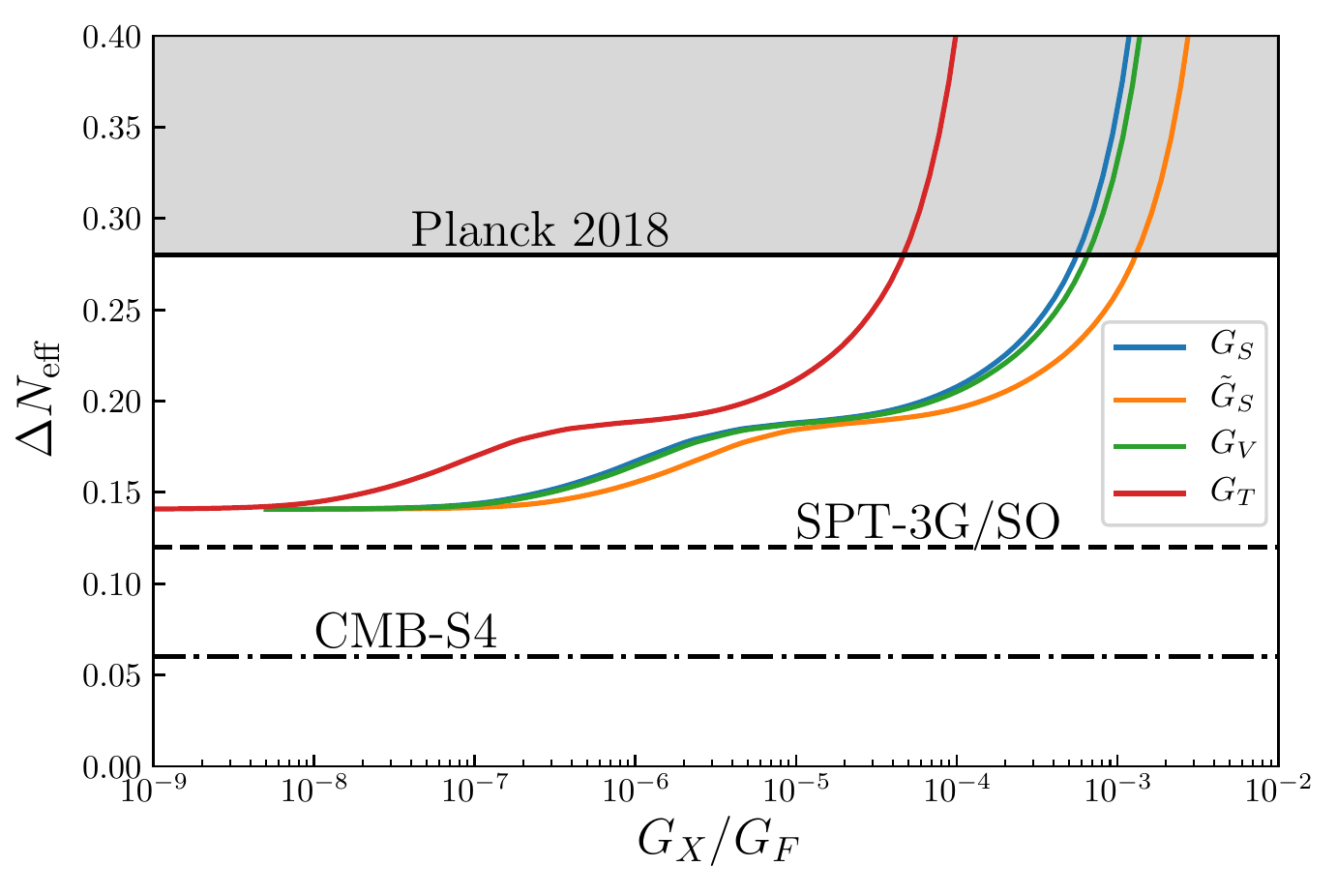}

\caption{Contributions of $\nu_{R}$ to $N_{{\rm eff}}$ in the presence of
$G_{S}$, $\tilde{G}_{S}$, $G_{V}$, $G_{T}$ interactions defined
in Eq.~(\ref{eq:b-9}). \label{fig:Neff} The experimental bounds
are presented at 2$\sigma$ (95\%) C.L.}
\end{figure}

\begin{figure}
\centering

\includegraphics[width=12cm]{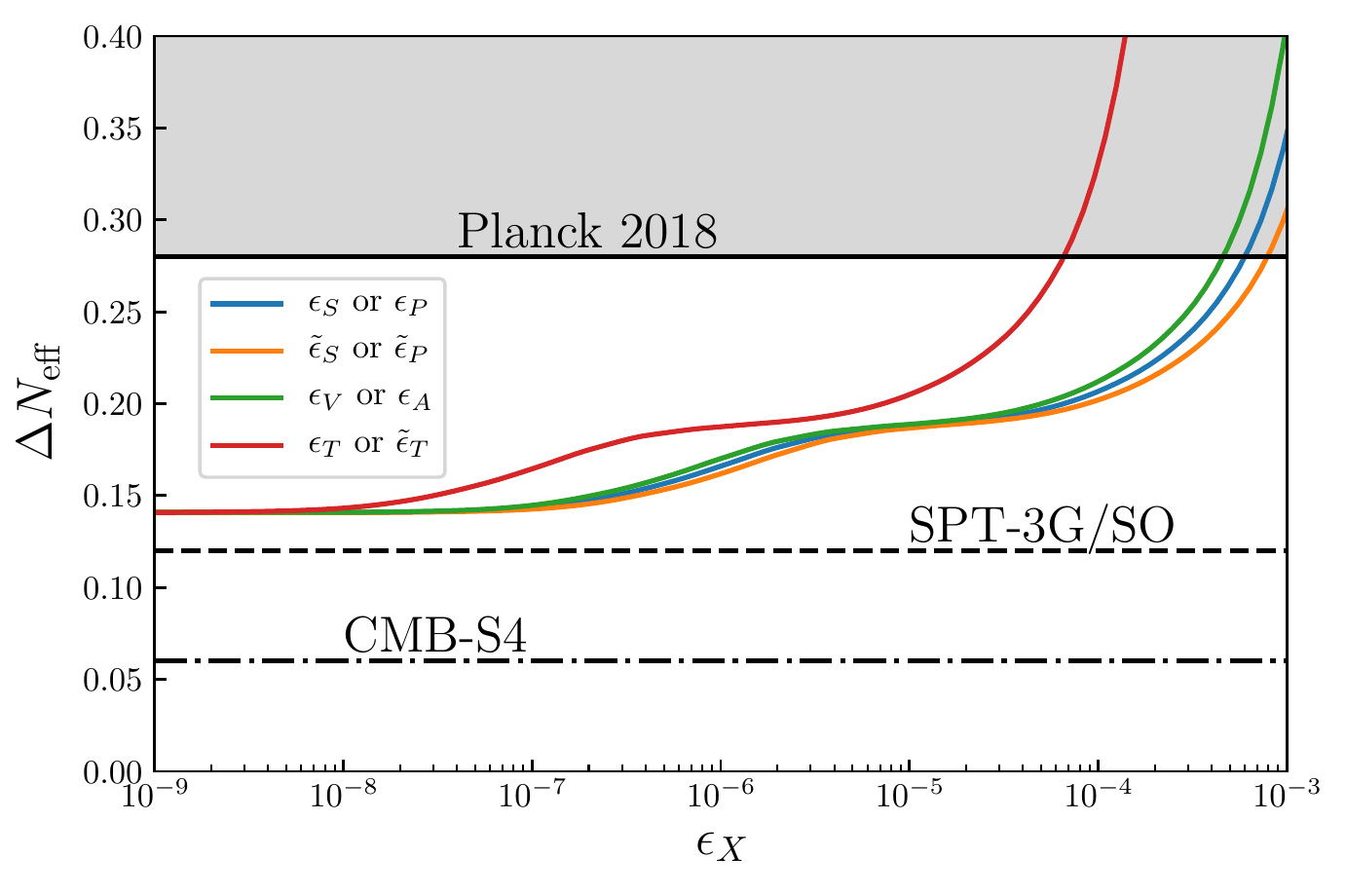}

\caption{Same as Fig.\ \ref{fig:Neff} for the $\epsilon_X$ from Eq.\ (\ref{eq:4f}). \label{fig:Neff_eps}}
\end{figure}

In Fig.~\ref{fig:Neff}, we compute $\Delta N_{{\rm eff}}$ for the
four different types of interactions ($G_{S}$, $\tilde{G}_{S}$,
$G_{V}$, $G_{T}$) and compare the results with current and future
experimental limits on $N_{{\rm eff}}$. Currently the Planck satellite~\cite{Akrami:2018vks,Aghanim:2018eyx}
has measured $N_{{\rm eff}}=2.99\pm0.17$ at 1$\sigma$ confidence
level (C.L.), which is so far the strongest limit on $N_{{\rm eff}}$.
We put a 2$\sigma$ bound (black solid curve) corresponding to $\Delta N_{{\rm eff}}<2.99+0.17\times2-3.045=0.285$.
 Future experiments such as the South Pole Telescope (SPT-3G)~\cite{Benson:2014qhw},
the Simons Observatory (SO)~\cite{Abitbol:2019nhf}, and CMB-S4~\cite{Abazajian:2016yjj,Abazajian:2019eic}
will significantly improve the measurement of $N_{{\rm eff}}$. The
SPT-3G is expected to be sensitive to $\Delta N_{{\rm eff}}$ larger
than 0.058 (1$\sigma$) and the SO sensitivity is very similar. So
we take $\Delta N_{{\rm eff}}<0.12$ as a $2\sigma$ limit for both
experiments. The CMB-S4 sensitivity is expected to reach 0.03 ($1$$\sigma$).
So we take $\Delta N_{{\rm eff}}<0.06$ at 2$\sigma$ C.L.\ for CMB-S4.

As shown in Fig.~\ref{fig:Neff}, the current limit on $\Delta N_{{\rm eff}}$
from the Planck 2018 data implies at $2\sigma$ the following upper limits on the effective coupling constants in Eqs.\ (\ref{eq:b-10})-(\ref{eq:b-13}): 
\begin{equation}
G_{S}<5.6\times10^{-4}\, G_{F},\ \tilde{G}_{S}<1.3\times10^{-3}\, G_{F},\ G_{V}<6.5\times10^{-4}\, G_{F},\ G_{T}<4.7\times10^{-5}\, G_{F}.\label{eq:d-27}
\end{equation}
In Fig.\ \ref{fig:Neff_eps} we show the result for the original parameters $\epsilon_X$ appearing in Eq.\ (\ref{eq:4f}). The limits at $2\sigma$ C.L.\ are 
\begin{equation}
\epsilon_{S,P}<5.9\times10^{-4},\ \tilde\epsilon_{S,P}<7.9\times10^{-4},\ \epsilon_{V,A}<4.5\times10^{-4},\
\epsilon_{T},\tilde\epsilon_{T}<6.5\times10^{-5}.
\end{equation}
Alternatively, we can get a feeling for the energy scale that is probed by evaluating 
$\sqrt{1/G_X}$, which would correspond to $m/g$, where $g$ is a new coupling and $m$ the mass of a mediator particle. This gives: 
  \begin{equation}
\sqrt{1/G_{S}} > 12.4 \, {\rm TeV},\
\sqrt{1/\tilde G_{S}} > 8.1 \, {\rm TeV},\
\sqrt{1/G_{V}} > 11.4 \, {\rm TeV},\
\sqrt{1/G_{T}} > 42.9 \, {\rm TeV}. \label{eq:d-28}
\end{equation}
Note that for the values shown in Figs.\ \ref{fig:T_R},  \ref{fig:Neff} and \ref{fig:Neff_eps}, it does not matter whether the initial temperature 
of the right-handed neutrinos is $T_{\rm SM}$ or zero. In the latter case $T_{\nu_R}$ approaches $T_{\rm SM}$ so quickly that no difference in the final result is visible. 
 
Future limits from the SPT-3G/SO and CMB-S4 experiments would be lower
than the minimal value ($\Delta N_{{\rm eff}}=0.14$) predicted in
this framework, which implies that the scenario considered in this
work could be fully excluded (up to some scenarios to be discussed below). 
On the other hand, if future measurements
find a nonzero $\Delta N_{{\rm eff}}$ larger than 0.14, Dirac neutrinos
with  BSM interactions would be one of the most well-motivated
scenarios to explain the deviation. \\

\begin{figure}
  \centering
  
  \includegraphics[width=12cm]{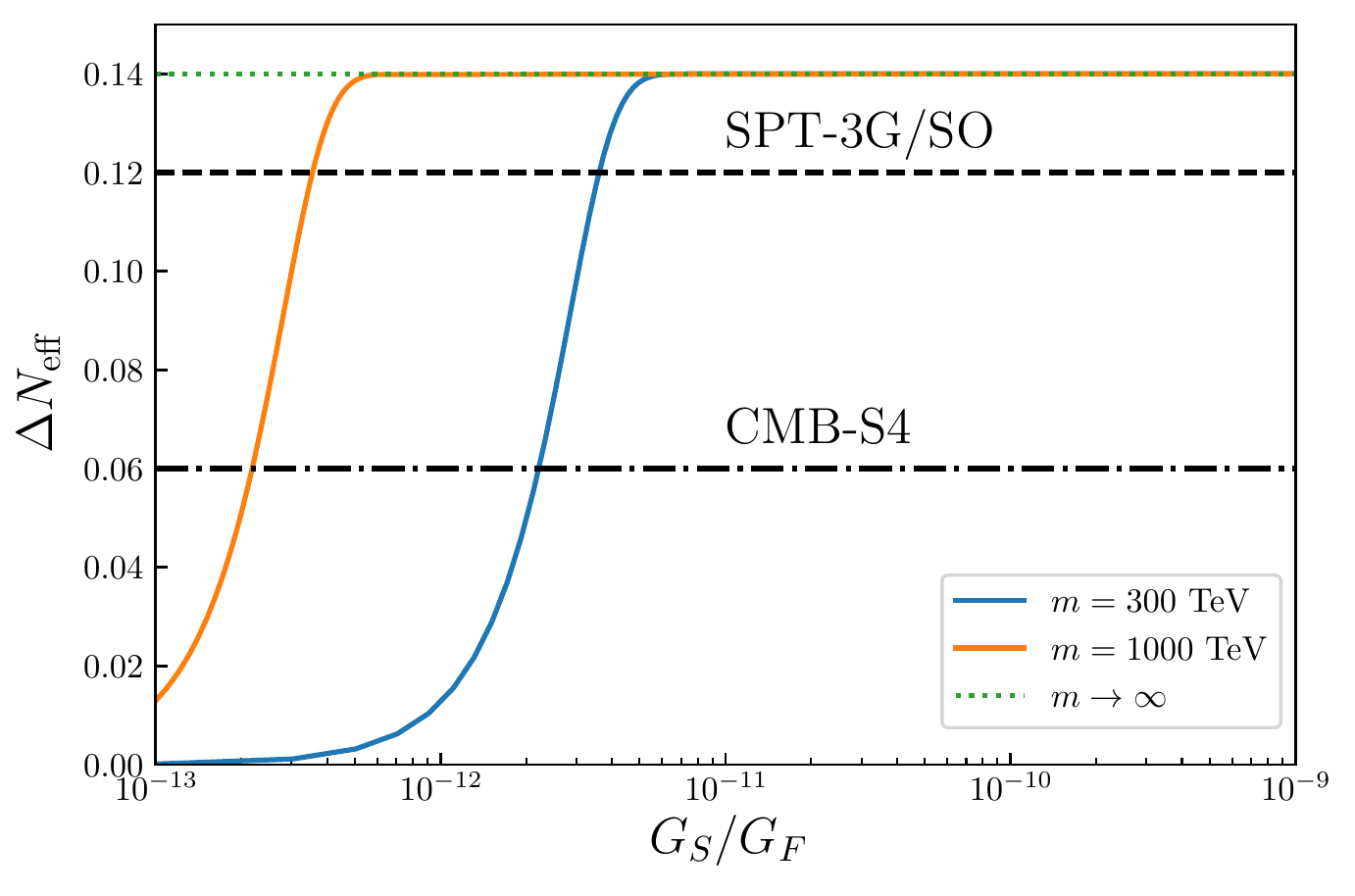}
  
  \caption{Illustration of possible modifications of the $\Delta N_{{\rm eff}}$-$G_S$ relation when $G_S=g^2/m^2$ decreases while $m$ is fixed at the given values. 
  \label{fig:Neff_m}
  }
\end{figure}

Note, however, that in Fig.~\ref{fig:Neff} one should not extrapolate the curves
to arbitrarily small $G_{X}$. 
This would lead to the conclusion that
even if $G_{X}\rightarrow0$, $\nu_{R}$ would still contribute to
$N_{{\rm eff}}$ with 0.14. Such an extrapolation relies on the assumptions
that $\nu_{R}$ had been in thermal equilibrium with the SM, and that the
number of effective degrees of freedom is indeed $g_{\star}^{(\rho)}=106.75$
for $T\gg1$ TeV. Actually none of these assumptions may hold for very small $G_{X}$ or very high $T$. Let us outline two  scenarios which would change the lower
bound of $\Delta N_{{\rm eff}}$. \\

(i) Consider that $G_{X}$ is mediated by a heavy boson:
\begin{equation}
G_{X}=\frac{g^{2}}{m^{2}},\label{eq:gm}
\end{equation}
where $g$ is a new coupling and $m$ is the mass of the new boson. For fixed $m$, $G_{X}\rightarrow0$
would imply $g\rightarrow0$. Let us examine Eq.~(\ref{eq:d-20})
in this limit. If the $\nu_{R}$ had ever been in thermal equilibrium, then for decreasing $G_{X}$, their decoupling temperature would increase according to Eq.~(\ref{eq:d-23}), and 
eventually would exceed $m$. 
Note that when $T\gg m$, the effective coupling would be $G_{X}\rightarrow g^{2}/T^{2}$, and hence the 
right-hand side of Eq.~(\ref{eq:d-20}) would be proportional to $g^{4}T$. On
the left-hand side, the Hubble rate $H\propto T^{2}$
increases faster than $g^{4}T$ as $T$ further rises. Therefore,
in this case, Eq.~(\ref{eq:d-20}) would have no solution with respect
to $T_{{\rm dec}}$, which implies that $\nu_{R}$ would never have been in thermal equilibrium 
with the SM. 

If one numerically solves
the Boltzmann equation, one can find that given an initial value  $T_{\nu_{R}}=0$, the temperature ratio $T_{\nu_{R}}/T_{{\rm SM}}$
 will eventually approach a constant ($<1$), which is known as
the freeze-in mechanism. Therefore, if $G_{X}\rightarrow0$ is interpreted
as $g\rightarrow0$ with $m$ fixed, $\Delta N_{{\rm eff}}$ should
vanish in this limit. In Fig.~\ref{fig:Neff_m}, we fix $m$ at two
 values and solve, using $G_S = g^2/(m^2 + T^2)$, the Boltzmann equation with initial $T_{\nu_{R}}=0$ 
to obtain the corresponding $\Delta N_{{\rm eff}}$ for varying $G_{S}$. 
As can be seen from Fig.~\ref{fig:Neff_m}, for $m$ fixed at finite
values, $\Delta N_{{\rm eff}}$ is suppressed for small $G_{S}$.  

(ii) The lower bound $\Delta N_{{\rm eff}}\geq0.14$ depends significantly on the
maximal value of $g_{\star}^{(s)}$. The maximal SM-value of $106.75$, would be changed
if there are more particles beyond the SM at higher energy scales. Note that 
from 2 MeV to 200 GeV, $g_{\star}^{(s)}$ increases by roughly
a factor of 10. It is possible that from the electroweak scale up
to the Planck scale, new physics substantially enhances
$g_{\star}^{(s)}$  by another factor of 10 or more. 
According to Eqs.~(\ref{eq:d-17}) 
and (\ref{eq:d-25}), taking $g_{\star}^{(s)}(T_{{\rm dec}})\approx10^{3}$
for example, one would have $\Delta N_{{\rm eff}}\approx3\times(10.75/1000)^{4/3}\approx0.007$,
which would be below the future CMB-S4 sensitivity.\\

As illustrated by the aforementioned two scenarios, the relic density of $\nu_{R}$ and 
$\Delta N_{{\rm eff}}$ may be suppressed by new physics or by 
the model-dependent UV-completions of the new interactions.  Here we refrain from further discussion and leave these possibilities to be studied in our future work. The bounds in Eq.\ (\ref{eq:d-27}) that we obtain on the new interactions can nevertheless be considered as robust.

\section{Conclusion\label{sec:Conclusion}}
\noindent
Dirac neutrinos are a particularly interesting case of simple and straightforward physics that influences the effective number of relativistic degrees of freedom ($N_{\rm eff}$) in the early Universe. Current and future data will put this possibility to the test. 

We have considered here Dirac neutrinos with their most general effective interactions, as formulated in Eq.~(\ref{eq:4f}),  and studied the constraints that Planck 2018 data puts on their strength. The new interactions would equilibrate the right-handed neutrinos and therefore lead to potentially large contributions to $N_{\rm eff}$. Confronting this with Planck 2018 data leads to limits on the effective interaction strength of the order of $10^{-3}$ to $10^{-5}$ in units of the Fermi constant (see Figs.~\ref{fig:Neff} and \ref{fig:Neff_eps}), or energy scales corresponding to up to 43 TeV and higher. Since the scenario of effective 4-$\nu$ operators predicts $\Delta N_{\rm eff}\geq 0.14$, future experiments such as CMB-S4 which is expected to reach a sensitivity of $\Delta N_{\rm eff}\sim 0.03$ can fully probe or exclude it. We commented on possibilities to avoid these conclusions.


\begin{acknowledgments}
\noindent
W.R.\ is supported by the DFG with grant RO 2516/7-1 in the Heisenberg program.
\end{acknowledgments}

\appendix

\section{Thermodynamics\label{sec:thermaldy}}

\noindent In this appendix, we briefly review some relevant aspects
of equilibrium thermodynamics which are used in this work.  Although
most of the formulae can be found in textbooks (e.g., Kolb \& Turner
\cite{Kolb}), we would like to address some subtle issues via this
brief review.

For particles in thermal equilibrium, their distribution $f$ obeys
the Bose-Einstein or Fermi-Dirac distributions:
\begin{equation}
f=\frac{1}{\exp\left(\frac{E-\mu}{T}\right)\mp1}\thinspace,\label{eq:b-3}
\end{equation}
where ``$\mp$'' is ``$-$'' for bosons and ``$+$'' for fermions.
The notations $E$, $T$, and $\mu$ are for the energy, temperature,
and chemical potential of the particles, respectively. The definitions
of energy density ($\rho$), number density ($n$), pressure ($P$),
and entropy density $(s)$ are
\begin{eqnarray}
\rho & \equiv & \int E\thinspace f(E)\thinspace\frac{g}{(2\pi)^{3}}d^{3}p\thinspace,\label{eq:b-4-1}\\
n & \equiv & \int f(E)\thinspace\frac{g}{(2\pi)^{3}}d^{3}p\thinspace,\label{eq:b-5-1}\\
P & \equiv & \int\frac{|\vec{p}|^{2}}{3E}f(E)\thinspace\frac{g}{(2\pi)^{3}}d^{3}p\thinspace,\label{eq:b-6-1}\\
s & \equiv & \frac{\rho+P}{T}\thinspace.\label{eq:b-7-1}
\end{eqnarray}
Here $g$ denotes the internal degrees of freedom. For massless or
relativistic particles with negligible $\mu$ (typically this implies
no particle-antiparticle asymmetry), all the above integrals can be
evaluated analytically:
\begin{eqnarray}
\rho & = & \frac{\pi^{2}}{30}gT^{4}\times\begin{cases}
1 & ({\rm boson})\\
7/8 & ({\rm fermion})
\end{cases}\thinspace,\label{eq:b-4-2}\\
n & = & \frac{\zeta(3)}{\pi^{2}}gT^{3}\times\begin{cases}
1 & ({\rm boson})\\
3/4 & ({\rm fermion})
\end{cases}\thinspace,\label{eq:b-5-2}\\
P & = & \rho/3\thinspace,\label{eq:b-6-2}\\
s & = & \frac{2\pi^{2}}{45}gT^{3}\times\begin{cases}
1 & ({\rm boson})\\
7/8 & ({\rm fermion})
\end{cases}\thinspace.\label{eq:b-7-2}
\end{eqnarray}
Here $\zeta(3)\approx1.202$ is a value of the Riemann zeta function.
For multiple species in thermal equilibrium with each other, it is
convenient to define effective degrees of freedom $g_{\star}^{(\rho)}$,
$g_{\star}^{(n)}$, $g_{\star}^{(P)}$, and $g_{\star}^{(s)}$ via
\begin{equation}
\rho=\frac{\pi^{2}}{30}g_{\star}^{(\rho)}T^{4},\ \ n=\frac{\zeta(3)}{\pi^{2}}g_{\star}^{(n)}T^{3},\ \ P=\frac{\pi^{2}}{90}g_{\star}^{(P)}T^{4},\ \ s=\frac{2\pi^{2}}{45}g_{\star}^{(s)}T^{3}.\label{eq:g_def}
\end{equation}
Here $g_{\star}^{(\rho)}$ is the most commonly used form of $g_{\star}$
in the literature, typically appearing without the superscript $(\rho)$.
For the SM, all these quantities have been comprehensively studied
and computed in Ref.~\cite{Husdal:2016haj}. Note that Eqs.~(\ref{eq:b-4-2})
to (\ref{eq:b-7-2}) hold only for relativistic particles while Eq.~(\ref{eq:g_def})
applies for both relativistic and non-relativistic particles. For
particles with arbitrary masses, one can always use Eqs.~(\ref{eq:b-4-1})
to (\ref{eq:b-7-1}) to compute $\rho$, $n$, $P$, and $s$, and
then compute the corresponding $g_{\star}^{(\rho)}$, $g_{\star}^{(n)}$,
$g_{\star}^{(P)}$ and $g_{\star}^{(s)}$ according to Eq.~(\ref{eq:g_def}).

When using Eq.~(\ref{eq:g_def}), it is important to note that $g_{\star}^{(\rho)}$,
$g_{\star}^{(n)}$, $g_{\star}^{(P)}$, and $g_{\star}^{(s)}$ are
also functions of $T$, which implies that in 
$d\rho/dT=4\rho/T+\rho/g_{\star}^{(\rho)} \, d g_{\star}^{(\rho)}\!/d T$,
the second term should not be ignored. It is also worth mentioning
that since the energy density and pressure are related by\footnote{See, e.g., Eq.~(3.67) in Ref.~\cite{Kolb}.}
\begin{equation}
dP=\frac{\rho+P}{T}dT,\label{eq:dp}
\end{equation}
one can derive an identity for $dg_{\star}^{(P)}/dT$, 
\begin{equation}
\frac{dg_{\star}^{(P)}}{dT}=3\thinspace\frac{g_{\star}^{(\rho)}-g_{\star}^{(P)}}{T},\label{eq:dg}
\end{equation}
which is technically useful to determine $g_{\star}^{(P)}(T)$ and
$g_{\star}^{(\rho)}(T)$ from each other  if only one of them is given.

When several particle species in the early Universe interact with
each other,  their distributions are governed by the Boltzmann equation~\cite{Dolgov:2002wy}:
\begin{equation}
\left[\frac{\partial}{\partial t}-H\thinspace\vec{p}\cdot\nabla_{\vec{p}}\right]f_{\psi}(\vec{p},\ t)=C_{\psi}^{(f)}.\label{eq:b-22}
\end{equation}
Here $\psi$ is a specific species of interest, and $f_{\psi}$ is
the distribution function of $\psi$, not necessarily in the form
of Eq.~(\ref{eq:b-3}) if $\psi$ is not in thermal equilibrium. The
right-hand side is a collision term which for a given process $\psi+a+b+\cdots\rightarrow i+j+\cdots$
 is computed by
\begin{eqnarray}
C_{\psi}^{(f)} & = & -\frac{1}{2E_{\psi}}\int d\Pi_{a}d\Pi_{b}\cdots d\Pi_{i}d\Pi_{j}\cdots(2\pi)^{4}\delta^{4}(p_{\psi}+p_{a}+p_{b}+\cdots-p_{i}-p_{j}-\cdots)\nonumber \\
 &  & \times S\left[|{\cal M}|_{\psi+a+b+\cdots\rightarrow i+j+\cdots}^{2}f_{\psi}f_{a}f_{b}\cdots(1\pm f_{i})(1\pm f_{j})\cdots\right.\nonumber \\
 &  & \ \ \ \ \left.-|{\cal M}|_{i+j+\cdots\rightarrow\psi+a+b+\cdots}^{2}f_{i}f_{j}\cdots(1\pm f_{\psi})(1\pm f_{a})(1\pm f_{b})\cdots\right],\label{eq:b-1-1}
\end{eqnarray}
with
\begin{equation}
d\Pi_{x}\equiv\frac{g_{x}}{(2\pi)^{3}}\frac{d^{3}p_{x}}{2E_{x}}\thinspace,\ \ x\in\{\psi,\ a,\ b,\ \cdots i,\ j,\ \cdots\}.\label{eq:b-2}
\end{equation}
Here ``$\pm$'' takes ``$+$'' for bosons or ``$-$'' for fermions;
 $S$ is a symmetry factor related to the number of identical particles
in the initial/final states, and ${\cal M}$ is the scattering amplitude
of the process specified in its subscript. 

Applying the $\vec{p}\cdot\nabla_{\vec{p}}$ operator in Eq.~(\ref{eq:b-22})
to the $f$ in Eq.~(\ref{eq:b-5-1}) gives 
\begin{equation}
\int\vec{p}\cdot\nabla_{\vec{p}}f\thinspace\frac{g}{(2\pi)^{3}}d^{3}p=\int\left[\nabla_{\vec{p}}(\vec{p}\thinspace f)-f\thinspace\nabla_{\vec{p}}\cdot\vec{p}\right]\thinspace\frac{g}{(2\pi)^{3}}d^{3}p=-3n\thinspace,\label{eq:b-23}
\end{equation}
where the $\nabla_{\vec{p}}(\vec{p}\thinspace f)$ term vanishes because
it is a total derivative (provided that $\vec{p}\thinspace f\rightarrow0$
if $p\rightarrow\infty$). Similarly, applying $\vec{p}\cdot\nabla_{\vec{p}}$
to the $f$ in Eq.~(\ref{eq:b-4-1}), we have 
\begin{equation}
\int E\vec{p}\cdot\nabla_{\vec{p}}f\thinspace\frac{g}{(2\pi)^{3}}d^{3}p=\int\left[\nabla_{\vec{p}}(E\vec{p}\thinspace f)-f\thinspace\nabla_{\vec{p}}\cdot(E\vec{p})\right]\thinspace\frac{g}{(2\pi)^{3}}d^{3}p=-3(\rho+P),\label{eq:b-24}
\end{equation}
where we have used $\nabla_{\vec{p}}E=\vec{p}/E$. From Eqs.~(\ref{eq:b-22}),
(\ref{eq:b-23}) and (\ref{eq:b-24}), we can obtain the following
integrated Boltzmann equations: 
\begin{eqnarray}
\frac{dn_{\psi}}{dt}+3Hn_{\psi} & = & C_{\psi}^{(n)},\label{eq:b}\\
\frac{d\rho_{\psi}}{dt}+3H(\rho+P) & = & C_{\psi}^{(\rho)},\label{eq:b-4}
\end{eqnarray}
where
\begin{eqnarray}
C_{\psi}^{(n)} & \equiv & \int C_{\psi}^{(f)}\thinspace\frac{g_{\psi}}{(2\pi)^{3}}d^{3}p_{\psi},\label{eq:Cn}\\
C_{\psi}^{(\rho)} & \equiv & \int C_{\psi}^{(f)}\thinspace\frac{E_{\psi}g_{\psi}}{(2\pi)^{3}}d^{3}p_{\psi}.\label{eq:Crho}
\end{eqnarray}
Here one may wonder about the symmetry factor $S$ in Eq.\ (\ref{eq:b-1-1}). In the
presence of identical particles, is the $S$ factor in the collision
term $C_{\psi}^{(f)}$  the same as the ones in $C_{\psi}^{(n)}$
and $C_{\psi}^{(\rho)}$? If among the particles $a,\ b, \ldots$
in Eq.~(\ref{eq:b-1-1}), $n$ of them are identical to $\psi$ and
other particles are not identical, then the $S$ factor should be
$\frac{1}{n!}$. However, when $d^{3}p_{\psi}$ further enters the
phase space integral in Eqs.~(\ref{eq:Cn}) or (\ref{eq:Crho}), the
number of identical particles in the phase space integral increases 
by one to $n+1$, hence leading to a factor of $\frac{1}{(n+1)!}$.
 On the other hand, when the $n$ identical particles happen to be
$\psi$, the process is $n$ times more efficient in the conversion
of particles or energy from $\psi$ to other particles. Therefore,
one should multiply the result by an additional factor of $1+n$.
Therefore, based on the number of identical particles in Eq.~(\ref{eq:b-1-1}),
the $S$ factor should be $\frac{1}{n!}$, while in Eqs.~(\ref{eq:Cn})
or (\ref{eq:Crho}) it should be $\frac{n+1}{(n+1)!}$, which is 
the same as that in Eq.~(\ref{eq:b-1-1}).

Another noteworthy issue concerns a potential difference in using
Eq.~(\ref{eq:b}) and (\ref{eq:b-4}). Consider an elastic scattering
process of $\psi$ with particles of another species $\psi'$: $\psi+\psi'\rightarrow\psi+\psi'$,
which eliminates one $\psi$ and produces another $\psi$  simultaneously.
The corresponding $C_{\psi}^{(n)}$ vanishes but $C_{\psi}^{(\rho)}\neq0$.
Although this process does not contribute to $\frac{dn_{\psi}}{dt}$
directly, it leads to energy conversion from $\psi'$ to $\psi$,
or vice versa. If each species keeps thermal equilibrium internally
and $T_{\psi}<T_{\psi'}$, then  the energy injected to $\psi$ via
this process will increase $\rho_{\psi}$ and $T_{\psi}$. Consequently,
$n_{\psi}$ has to be increased if $\psi$ is relativistic and the
internal thermal equilibrium of $\psi$ is maintained. This is usually
caused by self-interactions of $\psi$  which could lead to processes
such as $\psi+\overline{\psi}\rightarrow2\psi+2\overline{\psi}$.
Therefore, if $\psi$ keeps internal equilibrium via self-interactions,
the collision term in Eq.~(\ref{eq:b}) has to take into account such
processes, while in Eq.~(\ref{eq:b-4}) they can be ignored due to
energy conservation.

\section{Calculation of $|{\cal M}|^{2}$\label{sec:M2}}

\noindent In this appendix, we present the details of computing $|{\cal M}|^{2}$
for  the processes listed in Tab.~\ref{tab:t-ann-scat}.

Let us first start with the process 
\begin{equation}
\nu_{R}(p_{1})+\overline{\nu_{L}}(p_{2})\rightarrow\overline{\nu_{R}}(p_{3})+\nu_{L}(p_{4}).\label{eq:m-9}
\end{equation}
   Given the interactions in Eq.~(\ref{eq:b-9}), only $G_{S}$
and $G_{T}$ can lead to this process. In the presence of $G_{S}$
and $G_{T}$, the scattering amplitude reads:
\begin{eqnarray}
i{\cal M}^{s_{1}s_{2}s_{3}s_{4}}=\sum_{a=S,T} &  & \left\{ 2(iG_{a})\left[\overline{v_{2}}^{s_{2}}(p_{2})P_{R}\Gamma^{a}P_{R}u_{1}^{s_{1}}(p_{1})\right]\left[\overline{u_{4}}^{s_{4}}(p_{4})P_{R}\Gamma^{a}P_{R}v_{3}^{s_{3}}(p_{3})\right]\right.\nonumber \\
 &  & \left.-2(iG_{a})\left[\overline{u_{4}}^{s_{4}}(p_{4})P_{R}\Gamma^{a}P_{R}u_{1}^{s_{1}}(p_{1})\right]\left[\overline{v_{2}}^{s_{2}}(p_{2})P_{R}\Gamma^{a}P_{R}v_{3}^{s_{3}}(p_{3})\right]\right\} .\label{eq:m-3}
\end{eqnarray}
where $u_{1\cdots4}$ and $v_{1\cdots4}$ denote the external legs
of the particles and antiparticles in Eq.~(\ref{eq:m-9}); $p_{1\cdots4}$
are the corresponding momenta, $s_{1\cdots4}$ are the corresponding
spins. The factors of 2 in front of $iG_{a}$ arise because there 
are four different ways of assigning initial or final states to the
four $\nu$'s in each operator and two of them have the same amplitude.
The minus sign in the second row comes from exchanging fermion lines.

When computing $|{\cal M}_{(a)}|^{2}$, we sum over the spins of all
the initial and final states
\begin{equation}
|{\cal M}|^{2}=\sum_{s_{1}s_{3}}\sum_{s_{2}s_{4}}|{\cal M}^{s_{1}s_{2}s_{3}s_{4}}|^{2}.\label{eq:m-4}
\end{equation}
Note that for unpolarized scattering, typically there are factors
of $1/2$ in the spin summation. Here we are working on polarized scattering but the spins are still
summed over so that one can apply the trace technology. The difference is that here we do not have factors of $1/2$ in
the summation, provided that the amplitude in Eq.~(\ref{eq:m-3})
automatically vanishes if the spins do not match the projectors $P_{L}$
and $P_{R}$ in Eq.~(\ref{eq:m-3}).

If fully expanded, Eq.~(\ref{eq:m-3}) contains four terms and hence
Eq.~(\ref{eq:m-4}) contains 16 terms. The 16 terms can be converted
to either one trace of Dirac matrices, e.g.,
\begin{eqnarray}
 &  & \left[\overline{v_{2}}P_{R}\Gamma^{a}P_{R}u_{1}\right]\left[\overline{u_{4}}P_{R}\Gamma^{a}P_{R}v_{3}\right][\overline{u_{1}}P_{L}\Gamma^{b}P_{L}u_{4}][\overline{v_{3}}P_{L}\Gamma^{b}P_{L}v_{2}]\nonumber \\
 & \rightarrow & {\rm tr}\left[P_{R}\Gamma^{a}P_{R}u_{1}\overline{u_{1}}P_{L}\Gamma^{b}P_{L}u_{4}\overline{u_{4}}P_{R}\Gamma^{a}P_{R}v_{3}\overline{v_{3}}P_{L}\Gamma^{b}P_{L}v_{2}\overline{v_{2}}\right],\label{eq:m-10}
\end{eqnarray}
or two traces of two separate set of Dirac matrices, e.g.,
\begin{eqnarray}
 &  & \left[\overline{v_{2}}P_{R}\Gamma^{a}P_{R}u_{1}\right]\left[\overline{u_{4}}P_{R}\Gamma^{a}P_{R}v_{3}\right][\overline{u_{1}}P_{L}\Gamma^{b}P_{L}v_{2}][\overline{v_{3}}P_{L}\Gamma^{b}P_{L}u_{4}]\nonumber \\
 & \rightarrow & {\rm tr}\left[P_{R}\Gamma^{a}P_{R}u_{1}\overline{u_{1}}P_{L}\Gamma^{b}P_{L}v_{2}\overline{v_{2}}]{\rm tr}[P_{R}\Gamma^{a}P_{R}v_{3}\overline{v_{3}}P_{L}\Gamma^{b}P_{L}u_{4}\overline{u_{4}}\right].\label{eq:m-11}
\end{eqnarray}
Note that $\Gamma^{a}$ and $\Gamma^{b}$ should have different Lorentz
indices even if $a=b$. 

With the aforementioned details,  it is straightforward to  compute
$|{\cal M}|^{2}$:
\begin{equation}
|{\cal M}|^{2}=16|G_{S}-12G_{T}|^{2}(p_{1}\cdot p_{3})(p_{2}\cdot p_{4}).\label{eq:m-5}
\end{equation}
Now using crossing symmetry we can quickly obtain $|{\cal M}|^{2}$ for 
\begin{equation}
\nu_{R}(p_{1})+\nu_{R}(p_{2})\rightarrow\nu_{L}(p_{3})+\nu_{L}(p_{4})\label{eq:m-9-1}
\end{equation}
by replacing $p_{2}\rightarrow-p_{3}$ and $p_{3}\rightarrow-p_{2}$
in Eq.~(\ref{eq:m-5}):
\begin{equation}
|{\cal M}|^{2}=16|G_{S}-12G_{T}|^{2}(p_{1}\cdot p_{2})(p_{3}\cdot p_{4}).\label{eq:m-8}
\end{equation}

Next, let us consider the process 
\begin{equation}
\nu_{R}(p_{1})+\overline{\nu_{L}}(p_{2})\rightarrow\nu_{R}(p_{3})+\overline{\nu_{L}}(p_{4}),\label{eq:m-9-2}
\end{equation}
which can only be generated by $\tilde{G}_{S}$ and $G_{V}$. The
amplitude is simpler compared to the previous case because for each
operator there is only one way of assigning the initial/final states
to the 4 $\nu$'s in the operator:  
\begin{eqnarray*}
i{\cal M}^{s_{1}s_{2}s_{3}s_{4}} & = & i\tilde{G}_{S}\left[\overline{v_{2}}^{s_{2}}(p_{2})P_{R}P_{R}u_{1}^{s_{1}}(p_{1})\right]\left[\overline{u_{3}}^{s_{3}}(p_{3})P_{L}P_{L}v_{4}^{s_{4}}(p_{4})\right]\\
 &  & -iG_{V}\left[\overline{v_{2}}^{s_{2}}(p_{2})P_{R}\gamma^{\mu}P_{L}v_{4}^{s_{4}}(p_{4})\right]\left[\overline{u_{3}}^{s_{3}}(p_{3})P_{L}\gamma_{\mu}P_{R}u_{1}^{s_{1}}(p_{1})\right].
\end{eqnarray*}
Following a similar procedure, we obtain
\begin{equation}
|{\cal M}|^{2}=4|\tilde{G}_{S}-2G_{V}|^{2}(p_{1}\cdot p_{2})(p_{3}\cdot p_{4}).\label{eq:m-5-1}
\end{equation}
Again, using crossing symmetry, we can quickly obtain $|{\cal M}|^{2}$
for 
\begin{equation}
\nu_{R}(p_{1})+\overline{\nu_{R}}(p_{2})\rightarrow\nu_{L}(p_{3})+\overline{\nu_{L}}(p_{4})\label{eq:m-9-1-1}
\end{equation}
by replacing $p_{2}\rightarrow-p_{3}$ and $p_{3}\rightarrow-p_{2}$
in Eq.~(\ref{eq:m-5-1}):
\begin{equation}
|{\cal M}|^{2}=4|\tilde{G}_{S}-2G_{V}|^{2}(p_{1}\cdot p_{3})(p_{2}\cdot p_{4}).\label{eq:m-8-1}
\end{equation}
Finally, for the process 
\begin{equation}
\nu_{R}(p_{1})+\nu_{L}(p_{2})\rightarrow\nu_{R}(p_{3})+\nu_{L}(p_{4}),\label{eq:m-9-2-1}
\end{equation}
we replace $p_{2}\rightarrow-p_{4}$ and $p_{4}\rightarrow-p_{2}$
in Eq.~(\ref{eq:m-5-1}) and obtain:
\begin{equation}
|{\cal M}|^{2}=4|\tilde{G}_{S}-2G_{V}|^{2}(p_{1}\cdot p_{4})(p_{3}\cdot p_{2}).\label{eq:m-5-1-1}
\end{equation}

\section{Calculation of collision terms\label{sec:coll}}

\noindent In this appendix we present the calculation of collision
terms using the technique developed in Appendix A of Ref.~\cite{Dolgov:1997mb}.
To make the calculation applicable to both Eqs.~(\ref{eq:Cn}) and
(\ref{eq:Crho}), we focus on the following integral
\begin{equation}
C\equiv-\int d\Pi_{1}d\Pi_{2}d\Pi_{3}d\Pi_{4}(2\pi)^{4}\delta^{4}(p_{1}+p_{2}-p_{3}-p_{4})u(E_{1})F|{\cal M}|^{2},\label{eq:c-2}
\end{equation}
where $d\Pi_{i}\equiv\frac{1}{(2\pi)^{3}}\frac{d^{3}\mathbf{p_{i}}}{2E_{i}}$,
$u(E_{1})$ is a general function of $E_{1}$, and $F$ takes $F_{{\rm FD}}$
for Fermi-Dirac statistics or $F_{{\rm MB}}$ for Maxwell-Boltzmann
statistics, with $F_{{\rm FD}}$ and $F_{{\rm MB}}$ given as follows:
\begin{equation}
F_{{\rm FD}}=f_{1}f_{2}(1-f_{3})(1-f_{4})-f_{3}f_{4}(1-f_{1})(1-f_{2}),\ \ f_{i}=\frac{1}{\exp(E_{i}/T_{i})+1},\label{eq:c}
\end{equation}
\begin{equation}
F_{{\rm MB}}=f_{1}f_{2}-f_{3}f_{4},\ \ f_{i}=\frac{1}{\exp(E_{i}/T_{i})}.\label{eq:c-1}
\end{equation}
Here $T_{i}$ is the temperature of the $i$-th particle. The matrix
element squared $|{\cal M}|^{2}$ is in our work given as one of three different combination of 4-vector products, see Tab.\ \ref{tab:t}. We write it here in general as 
\begin{equation}
|{\cal M}|^{2}=G_{1}(p_{1}\cdot p_{2})(p_{3}\cdot p_{4})+G_{2}(p_{1}\cdot p_{3})(p_{2}\cdot p_{4})+G_{3}(p_{1}\cdot p_{4})(p_{2}\cdot p_{3}),\label{eq:c-5}
\end{equation}
where $G_{1}$, $G_{2}$, and $G_{3}$ are constants. 

When Eq.~(\ref{eq:c-2}) is applied to the collision term of number
density or energy density, i.e.\ $C^{(n)}$ in Eq.~(\ref{eq:Cn}) or
$C^{(\rho)}$ in Eq.~(\ref{eq:Crho}), we set $u(E_{1})=1$ or $u(E_{1})=E_{1}$,
respectively.

Using the identity
\begin{equation}
\delta^{3}(\mathbf{p_{1}}+\mathbf{p_{2}}-\mathbf{p_{3}}-\mathbf{p_{4}})=\int{e^{i(\mathbf{p_{1}}+\mathbf{p_{2}}-\mathbf{p_{3}}-\mathbf{p_{4}})\cdot\boldsymbol{\lambda}}\frac{d^{3}\boldsymbol{\lambda}}{(2\pi)^{3}}},\label{eq:c-6}
\end{equation}
we can  split Eq.~(\ref{eq:c-2}) into two integrals:
\begin{equation}
C=-\frac{1}{128\pi^{5}}\int\delta(E_{1}+E_{2}-E_{3}-E_{4})u(E_{1})F\thinspace D(p_{1},p_{2},p_{3},p_{4})\frac{p_{1}dp_{1}}{E_{1}}\frac{p_{2}dp_{2}}{E_{2}}\frac{p_{3}dp_{3}}{E_{3}}\frac{p_{4}dp_{4}}{E_{4}},\label{eq:c-4}
\end{equation}

\begin{equation}
D=\frac{p_{1}p_{2}p_{3}p_{4}}{256\pi^{6}}\int{d\Omega_{\lambda}}\int_{0}^{\infty}{\lambda^{2}d\lambda}\int{d\Omega_{1}e^{i\mathbf{p_{1}}\cdot\boldsymbol{\lambda}}}\int{d\Omega_{2}e^{i\mathbf{p_{2}}\cdot\boldsymbol{\lambda}}}\int{d\Omega_{3}e^{-i\mathbf{p_{3}}\cdot\boldsymbol{\lambda}}}\int{d\Omega_{4}e^{-i\mathbf{p_{4}}\cdot\boldsymbol{\lambda}}|{\cal M}|^{2}},\label{eq:c-3}
\end{equation}
where we have used spherical coordinates: $d^{3}\mathbf{p_{i}}=p_{i}^{2}dp_{i}d\Omega_{i}$
and $d^{3}\boldsymbol{\lambda}=\lambda^{2}d\lambda d\Omega_{\lambda}$. 
The integral $D$ can be analytically calculated given the general
form of $|{\cal M}|^{2}$ in Eq.~(\ref{eq:c-5}), as we shall work
out below.

In a Cartesian coordinate system with $\boldsymbol{\lambda}$ set
as the $z$-axis, we parameterize $\mathbf{p}_{i}$ as
\begin{equation}
\mathbf{p_{i}}=p_{i}(\sin\theta_{i}\cos\varphi_{i},\ \sin\theta_{i}\cos\varphi_{i},\ \cos\theta_{i}),\label{eq:c-26}
\end{equation}
so that
\begin{equation}
\mathbf{p_{i}}.\mathbf{p_{j}}=p_{i}p_{j}[\sin\theta_{i}\sin\theta_{i}\cos(\varphi_{i}-\varphi_{j})+\cos\theta_{i}\cos\theta_{j}],\label{eq:c-27}
\end{equation}
and 
\begin{equation}
{d\Omega_{i}e^{i\mathbf{p_{i}}\cdot\boldsymbol{\lambda}}}={d\cos\theta_{i}d\varphi_{i}e^{i\cos\theta_{i}p_{i}\lambda}}.\label{eq:c-28}
\end{equation}
For each term in Eq.~(\ref{eq:c-5}), it is straightforward to integrate
out $\varphi_{i}$ and $\theta_{i}$. Taking $|{\cal M}|^{2}\propto(p_{1}\cdot p_{2})(p_{3}\cdot p_{4})$
for example, we have
\begin{eqnarray}
 &  & \int{d\Omega_{1}e^{i\mathbf{p_{1}}\cdot\boldsymbol{\lambda}}}\int{d\Omega_{2}e^{i\mathbf{p_{2}}\cdot\boldsymbol{\lambda}}}\int{d\Omega_{3}e^{-i\mathbf{p_{3}}\cdot\boldsymbol{\lambda}}}\int{d\Omega_{4}e^{-i\mathbf{p_{4}}\cdot\boldsymbol{\lambda}}(p_{1}\cdot p_{2})(p_{3}\cdot p_{4})}\nonumber \\
 & = & \int dc_{1}d\varphi_{1}e^{ic_{1}p_{1}\lambda}dc_{2}d\varphi_{2}e^{ic_{2}p_{2}\lambda}\left[E_{1}E_{2}-p_{1}p_{2}s_{1}s_{2}\cos(\varphi_{1}-\varphi_{1})-p_{1}p_{2}c_{1}c_{2}\right]\nonumber \\
 &  & \times\{1\rightarrow3,\ 2\rightarrow4,\ p_{1}\rightarrow-p_{3},\ \ p_{2}\rightarrow-p_{4}\}\nonumber \\
 & = & \frac{16\pi^{2}}{\lambda^{2}p_{1}p_{2}}\left[E_{1}E_{2}S_{1}S_{2}+p_{1}p_{2}\left(C_{1}-\frac{S_{1}}{\lambda p_{1}}\right)\left(C_{2}-\frac{S_{2}}{\lambda p_{2}}\right)\right]\nonumber \\
 &  & \times\{1\rightarrow3,\ 2\rightarrow4,\ p_{1}\rightarrow-p_{3},\ \ p_{2}\rightarrow-p_{4}\},\label{eq:c-7}
\end{eqnarray}
where $(c_{i},\ s_{i})\equiv(\cos\theta_{i},\ \sin\theta_{i})$
and 
\begin{equation}
(C_{i},\ S_{i})\equiv(\cos\lambda p_{i},\ \sin\lambda p_{i}).\label{eq:c-9}
\end{equation}
Applying the above result to Eq.~(\ref{eq:c-3}), we obtain
\begin{eqnarray}
D^{(1\cdot2)(3\cdot4)} & = & \frac{4G_{1}}{\pi}\int_{0}^{\infty}{\lambda^{-2}d\lambda}\left[E_{1}E_{2}S_{1}S_{2}+p_{1}p_{2}\left(C_{1}-\frac{S_{1}}{\lambda p_{1}}\right)\left(C_{2}-\frac{S_{2}}{\lambda p_{2}}\right)\right]\nonumber \\
 &  & \times\left[E_{3}E_{4}S_{3}S_{4}+p_{3}p_{4}\left(C_{3}-\frac{S_{3}}{\lambda p_{3}}\right)\left(C_{4}-\frac{S_{4}}{\lambda p_{4}}\right)\right].\label{eq:c-8}
\end{eqnarray}
Here the superscript $(1\cdot2)(3\cdot4)$ is to remind us that so
far we have only considered the $G_{1}(p_{1}\cdot p_{2})(p_{3}\cdot p_{4})$
term.   Eq.~(\ref{eq:c-8}) can be decomposed as
\begin{eqnarray}
D^{(1\cdot2)(3\cdot4)} & = & G_{1}E_{1}E_{2}E_{3}E_{4}D_{SS}+G_{1}E_{1}E_{2}p_{3}p_{4}D_{SC}\nonumber \\
 &  & +G_{1}p_{1}p_{2}E_{3}E_{4}D_{CS}+G_{1}p_{1}p_{2}p_{3}p_{4}D_{CC},\label{eq:c-10}
\end{eqnarray}
where
\begin{eqnarray}
D_{SS} & = & \frac{4}{\pi}\int_{0}^{\infty}\frac{d\lambda}{\lambda^{2}}S_{1}S_{2}S_{3}S_{4},\label{eq:c-11}\\
D_{SC} & = & \frac{4}{\pi}\int_{0}^{\infty}\frac{d\lambda}{\lambda^{2}}S_{1}S_{2}\left(C_{3}-\frac{S_{3}}{\lambda p_{3}}\right)\left(C_{4}-\frac{S_{4}}{\lambda p_{4}}\right),\label{eq:c-12}\\
D_{CS} & = & \frac{4}{\pi}\int_{0}^{\infty}\frac{d\lambda}{\lambda^{2}}S_{3}S_{4}\left(C_{1}-\frac{S_{1}}{\lambda p_{1}}\right)\left(C_{2}-\frac{S_{2}}{\lambda p_{2}}\right),\label{eq:c-13}\\
D_{CC} & = & \frac{4}{\pi}\int_{0}^{\infty}\frac{d\lambda}{\lambda^{2}}\left(C_{1}-\frac{S_{1}}{\lambda p_{1}}\right)\left(C_{2}-\frac{S_{2}}{\lambda p_{2}}\right)\left(C_{3}-\frac{S_{3}}{\lambda p_{3}}\right)\left(C_{4}-\frac{S_{4}}{\lambda p_{4}}\right).\label{eq:c-14}
\end{eqnarray}
The integration of $\lambda$ in Eqs.~(\ref{eq:c-11})-(\ref{eq:c-14})
seems straightforward as one can express the trigonometric functions
to exponential functions and then convert the integrals to Euler's
gamma functions. It is worth mentioning, however, that one should
handle the branch cut singularities in the gamma functions carefully.
 Taking Eq.~(\ref{eq:c-11}) for example, we may meet integrals of
$\int\frac{d\lambda}{\lambda^{2}}\exp(i\lambda p)$, where $p$ can
be $p_{1}+p_{2}+p_{3}+p_{4}$, $p_{1}-p_{2}+p_{3}-p_{4}$, $p_{1}+p_{2}-p_{3}-p_{4}$,
etc. This integral is divergent but the divergence is expected to
be canceled out in Eq.~(\ref{eq:c-11}). One can regulate the integral
by limiting it in $\lambda\in[\epsilon,\ \infty)$ with $\epsilon>0$.
The result depends on whether $p>0$ or $<0$:
\begin{equation}
\lim_{\epsilon\rightarrow0^{+}}\int_{\epsilon}^{\infty}\frac{d\lambda}{\lambda^{2}}e^{i\lambda p}=\begin{cases}
\frac{1}{\epsilon}-\frac{\pi p}{2}-ip\left[\log(p\epsilon)+\gamma_{E}-1\right] & {\rm for\ }p>0\\
\frac{1}{\epsilon}+\frac{\pi p}{2}-ip\left[\log(-p\epsilon)+\gamma_{E}-1\right] & {\rm for}\ p<0
\end{cases},\label{eq:c-22}
\end{equation}
where $\gamma_{E}\approx0.577216$ is  Euler's constant. As a consequence,
the result of $D_{SS}$ depends on whether $p_{1}-p_{2}+p_{3}-p_{4}>0,$
$p_{1}+p_{2}-p_{3}-p_{4}>0$, $p_{1}-p_{2}-p_{3}+p_{4}>0,$ etc. 

With the above details being noted, we present the results of $D_{SS}$,
$D_{SC}$, $D_{CS}$ and $D_{CC}$: 
\begin{equation}
D_{SS}=\begin{cases}
\frac{1}{2}\left(-p_{1}+p_{2}+p_{3}+p_{4}\right) & ({\rm condition\ A})\\
p_{4} & ({\rm condition\ B})\\
\frac{1}{2}\left(p_{1}+p_{2}-p_{3}+p_{4}\right) & ({\rm condition\ C})\\
p_{2} & ({\rm condition\ D})
\end{cases},\label{eq:c-15}
\end{equation}
\begin{equation}
D_{SC}=\frac{1}{p_{3}p_{4}}\times\begin{cases}
\frac{\left(p_{1}-p_{2}\right){}^{3}-3\left(p_{3}^{2}+p_{4}^{2}\right)\left(p_{1}-p_{2}\right)+2\left(p_{3}^{3}+p_{4}^{3}\right)}{12} & ({\rm condition\ A})\\
\frac{p_{4}^{3}}{3} & ({\rm condition\ B})\\
\frac{-\left(p_{1}+p_{2}\right){}^{3}+3\left(p_{3}^{2}+p_{4}^{2}\right)\left(p_{1}+p_{2}\right)-2\left(p_{3}^{3}-p_{4}^{3}\right)}{12} & ({\rm condition\ C})\\
-\frac{1}{6}p_{2}\left(3p_{1}^{2}+p_{2}^{2}-3\left(p_{3}^{2}+p_{4}^{2}\right)\right) & ({\rm condition\ D})
\end{cases},\label{eq:c-16}
\end{equation}
\begin{equation}
D_{CS}=\left.D_{SC}\right|_{p_{1}\leftrightarrow p_{3},\ p_{2}\leftrightarrow p_{4}},\label{eq:c-29}
\end{equation}
\begin{equation}
D_{CC}=\frac{1}{p_{1}p_{2}p_{3}p_{4}}\times\begin{cases}
D_{CC}^{(\rm A)} & ({\rm condition\ A})\\
\frac{1}{30}\left(5\left(p_{1}^{2}+p_{2}^{2}+p_{3}^{2}\right)p_{4}^{3}-p_{4}^{5}\right) & ({\rm condition\ B})\\
D_{CC}^{(\rm C)} & ({\rm condition\ C})\\
\frac{1}{30}p_{2}^{3}\left(5p_{1}^{2}-p_{2}^{2}+5\left(p_{3}^{2}+p_{4}^{2}\right)\right) & ({\rm condition\ D})
\end{cases},\label{eq:c-15-1}
\end{equation}
\begin{eqnarray}
D_{CC}^{(\rm A)} & \equiv & \frac{p_{1}^{5}}{60}-\frac{1}{12}p_{2}^{2}p_{1}^{3}-\frac{1}{12}p_{3}^{2}p_{1}^{3}-\frac{1}{12}p_{4}^{2}p_{1}^{3}+\frac{1}{12}p_{2}^{3}p_{1}^{2}+\frac{1}{12}p_{3}^{3}p_{1}^{2}+\frac{1}{12}p_{4}^{3}p_{1}^{2}+\frac{1}{12}p_{2}^{2}p_{3}^{3}\nonumber \\
 &  & +\frac{1}{12}p_{2}^{2}p_{4}^{3}+\frac{1}{12}p_{3}^{2}p_{4}^{3}+\frac{1}{12}p_{2}^{3}p_{3}^{2}+\frac{1}{12}p_{2}^{3}p_{4}^{2}+\frac{1}{12}p_{3}^{3}p_{4}^{2}-\frac{p_{2}^{5}}{60}-\frac{p_{3}^{5}}{60}-\frac{p_{4}^{5}}{60},\label{eq:c-17}
\end{eqnarray}
\[
D_{CC}^{(\rm C)}=\left.D_{CC}^{(\rm A)}\right|_{p_{1}\leftrightarrow p_{3},\ p_{2}\leftrightarrow p_{4}}.
\]
Here we need to distinguish four conditions: 
\begin{eqnarray}
({\rm condition\ A}): & \  & p_{1}+p_{2}\geq p_{3}+p_{4}\land p_{1}+p_{4}\geq p_{2}+p_{3}\land p_{1}\geq p_{2}\land p_{3}\geq p_{4},\label{eq:c-18}\\
({\rm condition\ B}): & \  & p_{1}+p_{2}\geq p_{3}+p_{4}\land p_{1}+p_{4}<p_{2}+p_{3}\land p_{1}\geq p_{2}\land p_{3}\geq p_{4},\label{eq:c-19}\\
({\rm condition\ C}): & \  & p_{1}+p_{2}<p_{3}+p_{4}\land p_{1}+p_{4}<p_{2}+p_{3}\land p_{1}\geq p_{2}\land p_{3}\geq p_{4},\label{eq:c-20}\\
({\rm condition\ D}): & \  & p_{1}+p_{2}<p_{3}+p_{4}\land p_{1}+p_{4}\geq p_{2}+p_{3}\land p_{1}\geq p_{2}\land p_{3}\geq p_{4}.\label{eq:c-21}
\end{eqnarray}
Note that here we only present results for $p_{1}\geq p_{2}\land p_{3}\geq p_{4}$.
Since Eqs.~(\ref{eq:c-11})-(\ref{eq:c-14}) are symmetric under $1\leftrightarrow2$
and (or) $3\leftrightarrow4$, results for other possibilities such
as $p_{1}\geq p_{2}\land p_{3}<p_{4}$,  $p_{1}<p_{2}\land p_{3}\geq p_{4}$,
and $p_{1}<p_{2}\land p_{3}<p_{4}$ can be obtained by exchanging
$1\leftrightarrow2$ and (or) $3\leftrightarrow4$. 

So far we have not made any assumptions about the particle masses,
so the above calculations apply to both massless and massive particles.

Next, we proceed with the integral in Eq.~(\ref{eq:c-4}). Taking
the massless assumption $E_{i}=p_{i}$, Eq.~(\ref{eq:c-4}) can be
written as 
\begin{equation}
C^{(1\cdot2)(3\cdot4)}=-\frac{G_{1}}{128\pi^{5}}\int u(p_{1})F\thinspace p_{1}p_{2}p_{3}p_{4}(D_{SS}+D_{SC}+D_{CS}+D_{CC})dp_{1}dp_{3}dp_{4},\label{eq:c-23}
\end{equation}
where $p_{2}$ should be replaced by $p_{3}+p_{4}-p_{1}$. Using the
$D$-functions in Eqs.~(\ref{eq:c-15})-(\ref{eq:c-15-1}) and $F=F_{{\rm MB}}$
in Eq.~(\ref{eq:c-1}),  we obtain 
\begin{equation}
C^{(1\cdot2)(3\cdot4)}=\frac{G_{1}}{8\pi^{5}}\times\begin{cases}
3T_{3}^{4}T_{4}^{4}-3T_{1}^{4}T_{2}^{4} & {\rm for\ number\ density}\\
6T_{3}^{4}T_{4}^{4}\left(T_{3}+T_{4}\right)-12T_{1}^{5}T_{2}^{4} & {\rm for\ energy\ density}
\end{cases}.\label{eq:c-24}
\end{equation}
Here the superscript $(1\cdot2)(3\cdot4)$ reminds us that the result
is only for the $G_{1}(p_{1}\cdot p_{2})(p_{3}\cdot p_{4})$ term.
For other two terms in $|{\cal M}|^{2}$, namely $G_{2}(p_{1}\cdot p_{3})(p_{2}\cdot p_{4})$
and $G_{3}(p_{1}\cdot p_{4})(p_{2}\cdot p_{3})$, the calculation
is similar and we find that Eq.~(\ref{eq:c-23}) should be modified
to:
\begin{equation}
C^{(1\cdot3)(2\cdot4)}=-\frac{G_{2}}{128\pi^{5}}\int u(p_{1})F\thinspace p_{1}p_{2}p_{3}p_{4}\left[D_{SS}-D_{SC}-D_{CS}+D_{CC}\right]_{p_{2}\leftrightarrow p_{3}}dp_{1}dp_{3}dp_{4},\label{eq:c-23-1}
\end{equation}
and 
\begin{equation}
C^{(1\cdot4)(2\cdot3)}=-\frac{G_{3}}{128\pi^{5}}\int u(p_{1})F\thinspace p_{1}p_{2}p_{3}p_{4}\left[D_{SS}-D_{SC}-D_{CS}+D_{CC}\right]_{p_{2}\leftrightarrow p_{4}}dp_{1}dp_{3}dp_{4}.\label{eq:c-23-2}
\end{equation}
Note that the minus signs before $D_{SC}$ and $D_{CS}$ originate
from the minus signs in $e^{-i\mathbf{p_{3}}\cdot\boldsymbol{\lambda}}$
and $e^{-i\mathbf{p_{4}}\cdot\boldsymbol{\lambda}}$.  
The results of Eqs.~(\ref{eq:c-23-1}) and (\ref{eq:c-23-2}) read:
\begin{equation}
C^{(1\cdot3)(2\cdot4)}=\frac{G_{2}}{8\pi^{5}}\times\begin{cases}
T_{3}^{4}T_{4}^{4}-T_{1}^{4}T_{2}^{4} & {\rm for\ number\ density}\\
T_{3}^{4}T_{4}^{4}\left(T_{3}+3T_{4}\right)-4T_{1}^{5}T_{2}^{4} & {\rm for\ energy\ density}
\end{cases},\label{eq:c-24-1}
\end{equation}
\begin{equation}
C^{(1\cdot4)(2\cdot3)}=\frac{G_{3}}{8\pi^{5}}\times\begin{cases}
T_{3}^{4}T_{4}^{4}-T_{1}^{4}T_{2}^{4} & {\rm for\ number\ density}\\
T_{3}^{4}T_{4}^{4}\left(3T_{3}+T_{4}\right)-4T_{1}^{5}T_{2}^{4} & {\rm for\ energy\ density}
\end{cases}.\label{eq:c-24-2}
\end{equation}
In practical use, we often  have: (i) $T_{1}=T_{2},\ T_{3}=T_{4}$,
(ii) $T_{1}=T_{3},\ T_{2}=T_{4}$, or (iii) $T_{1}=T_{4},\ T_{2}=T_{3}$.
Case (i) appears when computing the collision term of an annihilation
process, and the last two cases apply to $\nu_{R}$ scattering with
$\nu_{L}$. In Tab.~\ref{tab:t-ann-scat}, we summarize the results
of the collision terms for cases (i) and (ii). For case (iii), the
result can be obtained from (ii) with $T_{3}\leftrightarrow T_{4}$
and $p_{3}\leftrightarrow p_{4}$. 

The analytical results in Eqs.~(\ref{eq:c-24}), (\ref{eq:c-24-1}),
and (\ref{eq:c-24-2}) are only for Maxwell-Boltzmann statistics.
For Fermi-Dirac statistics,  we numerically evaluate Eqs.~(\ref{eq:c-23}),
(\ref{eq:c-23-1}), and (\ref{eq:c-23-2}) with $F=F_{{\rm FD}}$
given in Eq.~(\ref{eq:c}). Then we compute the ratio between the
Fermi-Dirac result ($C_{{\rm FD}}$) and the Maxwell-Boltzmann one
($C_{{\rm MB}}$):
\begin{equation}
1-\delta_{{\rm FD}}\equiv\frac{C_{{\rm FD}}}{C_{{\rm MB}}}.\label{eq:c-25}
\end{equation}
 The ratio is temperature-dependent. But for the aforementioned three
cases (i, ii, iii), $1-\delta_{{\rm FD}}$ only depends on $\Delta T/T_{1}$
where $\Delta T$ is the difference between $T_{1}$ and the other
different temperature. When $\Delta T/T_{1}$ is large, the collision
term is not important because it implies that $\nu_{R}$ has decoupled.
So we are mainly interested in the value of $1-\delta_{{\rm FD}}$
when $\Delta T/T_{1}\ll1$ and we have found that in this case $1-\delta_{{\rm FD}}$
is insensitive to $\Delta T/T_{1}$. We compute the values of $1-\delta_{{\rm FD}}$
in the limit $\Delta T/T_{1}\rightarrow0$, and the results are summarized
in Tab.~\ref{tab:t-ann-scat}.

\begin{table*}
\caption{\label{tab:t-ann-scat}Collision terms for energy density ($C^{(\rho)}$)
and for number density ($C^{(n)}$). The analytical expressions for
$C_{{\rm MB}}^{(\rho)}$ and $C_{{\rm MB}}^{(n)}$ have been computed
assuming  Maxwell-Boltzmann statistics. For Fermi-Dirac statistics,
one should multiply $C_{{\rm MB}}$ by the numerical factors $1-\delta_{{\rm FD}}$
to include the difference.}

\begin{ruledtabular}
\begin{tabular}{cc|cccc|ccc}
 & $|M^{2}|$  & \multicolumn{4}{c|}{$T_{1}=T_{2}$, $T_{3}=T_{4}$ (annihilation)} & \multicolumn{3}{c}{$T_{1}=T_{3}$, $T_{2}=T_{4}$ (scattering)}\tabularnewline
\hline 
 &  & $C_{{\rm MB}}^{(\rho)}$  & $1-\delta_{{\rm FD}}^{(\rho)}$  & $C_{{\rm MB}}^{(n)}$  & $1-\delta_{{\rm FD}}^{(n)}$ & $C_{{\rm MB}}^{(\rho)}$  & $1-\delta_{{\rm FD}}^{(\rho)}$  & $C_{{\rm {\rm MB}}}^{(n)}$ \tabularnewline
\rule[-2ex]{0pt}{6ex} & $(p_{1}\cdot p_{2})(p_{3}\cdot p_{4})$  & $\frac{3(T_{3}^{9}-T_{1}^{9})}{2\pi^{5}}$  & 0.8840  & $\frac{3(T_{3}^{8}-T_{1}^{8})}{8\pi^{5}}$  & 0.8521 & $\frac{3T_{2}^{4}T_{1}^{4}(T_{2}-T_{1})}{4\pi^{5}}$  & 0.8249  & 0 \tabularnewline
\rule[-2.5ex]{0pt}{6ex}  &  $(p_{1}\cdot p_{3})(p_{2}\cdot p_{4})$  & $\frac{T_{3}^{9}-T_{1}^{9}}{2\pi^{5}}$  & 0.8841  & $\frac{T_{3}^{8}-T_{1}^{8}}{8\pi^{5}}$ & 0.8523 & $\frac{3T_{2}^{4}T_{1}^{4}(T_{2}-T_{1})}{8\pi^{5}}$  & 0.8118  & 0\tabularnewline
 & $(p_{1}\cdot p_{4})(p_{2}\cdot p_{3})$  & $\frac{T_{3}^{9}-T_{1}^{9}}{2\pi^{5}}$  & 0.8841  & $\frac{T_{3}^{8}-T_{1}^{8}}{8\pi^{5}}$ & 0.8523  & $\frac{T_{2}^{4}T_{1}^{4}(T_{2}-T_{1})}{8\pi^{5}}$  & 0.8518  & 0\tabularnewline
\end{tabular}\end{ruledtabular}

\end{table*}

\bibliographystyle{JHEP}
\bibliography{ref}

\end{document}